\documentclass[a4paper,11pt]{article}
\pdfoutput=1 % if your are submitting a pdflatex (i.e. if you have
\usepackage[english]{babel}
\usepackage{microtype,xspace}
\usepackage[T2A]{fontenc}
\usepackage[utf8]{inputenc}	% Allows for writing special characters in the tex-file

\usepackage{jheppub} % for details on the use of the package, please
\usepackage{bm,amssymb,amsmath,amsfonts,slashed,graphicx,multirow,soul,mathtools,xspace,array,comment,color}
\hypersetup{colorlinks=true, allcolors=blue}  % set hyperref options here
\usepackage{adjustbox}
\usepackage{siunitx}
\usepackage{bm} % allows to use \bm{} to make bold math symbols
\usepackage{float}
\allowdisplaybreaks
\usepackage{ bbold }
\usepackage{subfigure}
\usepackage[normalem]{ulem} % for crossing out text
\usepackage{enumerate}
\usepackage{multirow}
\usepackage{booktabs}
\usepackage{xparse}
\usepackage{etoolbox}
\usepackage{physics}
\usepackage{hypcap}
\usepackage{siunitx}
\usepackage{xcolor}
\usepackage{comment}
\usepackage[capitalise]{cleveref}

\newcommand{\BFmu}{{\bf \mu}}
\newcommand{\BFmubar}{{\bf \bar{\mu}}}

\def\be{\begin{equation}}
\def\ee{\end{equation}}
\newcommand{\TeV}{\text{TeV}}
\newcommand{\GeV}{\text{GeV}}

\arxivnumber{} % if you have one

\title{\boldmath Testing the Neutrino Content of the Muon\\ at Muon Colliders}

\author[1]{Rodolfo Capdevilla,}
\author[2,3,4]{Francesco Garosi,}
\author[4]{David Marzocca,}
\author[5,6]{Bernd Stechauner}

\affiliation[1]{Particle Theory Department, Fermi National Accelerator Laboratory, Batavia, IL 60510, USA}
\affiliation[2]{Max-Planck-Institute f\"ur Physik, Boltzmannstra\ss e 8, 85748 Garching, Germany}
\affiliation[3]{SISSA International School for Advanced Studies, Via Bonomea 265, 34136, Trieste, Italy}
\affiliation[4]{INFN, Sezione di Trieste, SISSA, Via Bonomea 265, 34136, Trieste, Italy}
\affiliation[5]{CERN, Espl. des Particules 1, 1211 Geneva 23, Switzerland}
\affiliation[6]{TU Wien, Karlsplatz 13, 1040 Vienna, Austria}

\abstract{
Collinear emission of $W$ bosons off a high-energy muon induces a large muon-neutrino component among the Parton Distribution Functions (PDFs) of a muon. In this paper we study the phenomenology related to the $\nu_\mu$ PDF at future high-energy muon colliders.
We examine total rates and differential distributions of the $e \bar{\nu}_e$ and $W \gamma$ production processes, which receive a large, and often dominant, contribution from this PDF, allowing for a detailed experimental study.
As a demonstration of the impact the $\nu_\mu$ PDF could have for searches of new physics, we study the charged-current pair production of a couple of heavy states, components of a ${\rm SU}(2)_L$ doublet.
In both $e\bar{\nu}_e$ production and charged-current pair production of heavy states, we compare results obtained using PDFs with those of a fixed-order simulation.
}

\begin{document}
\preprint{FERMILAB-PUB-24-0575-T}
\maketitle
\flushbottom

%\begin{fmffile}{marrows}

%-------------------------------------------------
%-------------------------------------------------
\section{ Introduction}
\label{sec:intro}

The Standard Model (SM) of particle physics offers our best understanding of the physics of fundamental particles and their interactions. It provides a plethora of precise predictions that have been confirmed at particle accelerators over many decades, culminating with the discovery of the Higgs boson in 2012 by the Large Hadron Collider (LHC).
Nevertheless, fundamental physics at multi-TeV energy scales remains relatively untested and the presence of physics beyond the SM remains an exciting possibility, well motivated by the instability of the electroweak scale under quantum fluctuations.
Exploring this energy range is the main scope of the next generation of colliders, be it indirectly with very precise measurements of electroweak (EW) and Higgs processes or directly by reaching multi-TeV center-of-mass partonic energies in direct searches. A muon collider (MuC) offers the unique opportunity to
explore both these avenues with the same machine~\cite{Long:2020wfp,AlAli:2021let,Aime:2022flm,Black:2022cth,Accettura:2023ked,Andreetto2024,Accettura:2024qnk} (see also \cite{
Han:2020pif,Forslund:2022xjq,Ruhdorfer:2023uea,Forslund:2023reu,Andreetto:2024rra,Li:2024joa,
Han:2020uak,Buttazzo:2020uzc,Capdevilla:2021fmj,Bottaro:2021snn,Chen:2022msz,Bottaro:2022one,Azatov:2022itm,Liu:2023jta,Korshynska:2024suh,Capdevilla:2024bwt}).

Independently of the possible presence of new physics, SM dynamics has never been observed at these large energies and it offers a large number of new phenomena to be studied both experimentally and theoretically.
Specifically, at energies much larger than the EW scale the effects of EW masses become negligible and EW symmetry becomes effectively restored. The phenomena related to EW restoration are multiple and offer a very interesting physics program with assured deliverables for future high-energy colliders. They encompass effects such as the onset of large EW Sudakov double-logarithmic corrections \cite{Amati:1980ch,Ciafaloni:1998xg,Ciafaloni:2000gm,Ciafaloni:2000df,Ciafaloni:2000rp,Ciafaloni:2001vt,Manohar:2018kfx}, collinear emission of EW radiation \cite{Chen:2016wkt}, and EW PDFs \cite{Dawson:1984gx,Kane:1984bb,Ciafaloni:2001mu,Ciafaloni:2005fm,Bauer:2017isx,Bauer:2017bnh,Bauer:2018arx,Han:2020uid,Han:2021kes,Garosi:2023bvq,Ciafaloni:2024alq}. 
While some of these effects are already relevant at the LHC and will become even more so at a more energetic hadron collider, they play a central role in the physics of a MuC, since the effects of QCD interactions are suppressed by the non-colored nature of the muon.

The effects of initial-state radiation (ISR), that give rise to PDFs, are well known and have been widely studied at electron-positron colliders. 
The emission of copious amount of collinear photons \cite{Fermi:1924tc,Landau:1934zj,vonWeizsacker:1934nji,Williams:1934ad} causes a reduction of the viable center of mass (CoM) energy for the collisions and an increase of the probability for gamma-gamma collisions.
When the energy of the lepton increases and EW restoration becomes manifest, QED interactions must be substituted by the complete SM and the collinear emission of $W$ and $Z$ bosons becomes important.
In this sense, a multi-TeV lepton collider becomes effectively a {\it gauge boson collider} \cite{Costantini:2020stv,Ruiz:2021tdt}.
A peculiar feature of EW radiation, not present in the cases of QED or QCD, is that the valence lepton can transform into a neutrino via the emission of a collinear $W$ boson, contributing to the neutrino component of the lepton's PDFs.
Being dominated by the emission of low-energy $W$ bosons, the neutrinos generated in this way typically carry a large fraction of the original lepton energy. The neutrino PDF thus grows at large momentum fractions $x\simeq 1$, where it dominates over the gauge bosons PDFs \cite{Han:2020uid,Han:2021kes,Garosi:2023bvq}.

In this paper, we investigate several aspects of the phenomenology of muon neutrino PDFs at a future MuC.
We start in \cref{sec:nuPDF} by reviewing the neutrino PDF and providing an approximate analytic expression to the full numerical result, which we take from Ref.~\cite{Garosi:2023bvq}.
In \cref{sec:measurement} we study two processes where the contribution from the neutrino PDF is dominant and therefore could offer a potential cross-check for theoretical predictions. Specifically, we focus on $\mu \bar{\mu} \to e \nu$ and $\mu \bar{\mu} \to W \gamma$, where $\mu$ ($\bar\mu$) represents the muon (antimuon) beam, including its full parton content. We calculate total and differential rates to assess the sensitivity to the contribution of the neutrino PDF over the background from Vector Boson Fusion (VBF) as a function of the $p_T$ and rapidity of the final-state particles. We also compare the PDF results with those obtained at leading order (LO) with {\tt MadGraph5\_aMC} \cite{Alwall:2014hca} (in the following, MG5).
In \cref{sec:CCBSM} we study an application of neutrino PDFs for searches of physics beyond the SM. In case of pair production of heavy states, that typically proceeds via $\mu^- \mu^+$ annihilation in neutral current, the neutrino PDF opens the possibility to study charged-current production as well, e.g. $\nu_\mu \mu^+ \to W^* \to X_1 X_2^\dagger$, providing additional tests of new physics scenarios. We study the example of a heavy scalar ${\rm SU}(2)_L$ doublet, comparing the production cross section obtained using the neutrino PDF with the one derived from a LO MG5 simulation.
Finally, we conclude in \cref{sec:conclusions}.

%-------------------------------------------------
%-------------------------------------------------
\section{The muon neutrino PDF of a muon}
\label{sec:nuPDF}

When the transverse momentum of a splitting process in the initial-state radiation is much smaller than the typical energy of the subsequent hard scattering, i.e. in the collinear limit $p_T^{\rm ISR} \ll E_{\rm hard}$, then it is possible to factorize the splitting amplitude from the hard scattering one, up to small power corrections \cite{Kunszt:1987tk,Borel:2012by,Cuomo:2019siu}. This results in the well-known formalism of parton distribution functions, that can be used to describe the resummed multiple emission of collinear initial-state radiation: $f_i(x,Q^2)$ describes the probability of finding the parton $i$ inside the original particle, carrying a longitudinal momentum fraction $x$ and at a factorization scale $Q$.
In the case of lepton colliders, PDFs can be derived analytically from first principles by solving differential DGLAP equations~\cite{Gribov:1972ri,Dokshitzer:1977sg,Altarelli:1977zs}. At leading order, the boundary condition is set by imposing that at a factorization scale equal to the lepton mass, the only non-vanishing PDF is the one of the valence lepton itself and is a Dirac delta: $f_{\ell_{\rm val}}^{(\alpha^0)}(x, m^2_{\ell_{\rm val}}) = \delta(1-x)$.

At the lowest order in QED, $\mathcal{O}(\alpha)$, the only splitting process is $\ell^- \to \ell^- \gamma$, which generates a contribution to the photon PDF of the lepton, known at LO since a long time as the effective photon approximation \cite{Fermi:1924tc,vonWeizsacker:1934nji,Williams:1934ad,Landau:1934zj}, as well as a correction to the zeroth-order PDF of the valence lepton. This correction presents a soft IR divergence when the energy of the emitted photon goes to zero, which, in the case of QED, is regulated by loop corrections to the electron self-energy, that must be included at the same order.

\begin{figure}[t]
\centering
    \includegraphics[width=0.48\textwidth]{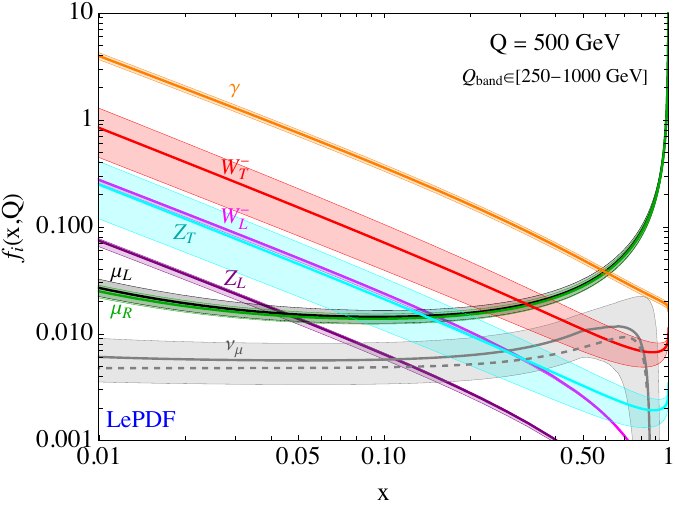}
    %\hfill
    \includegraphics[width=0.48\textwidth]{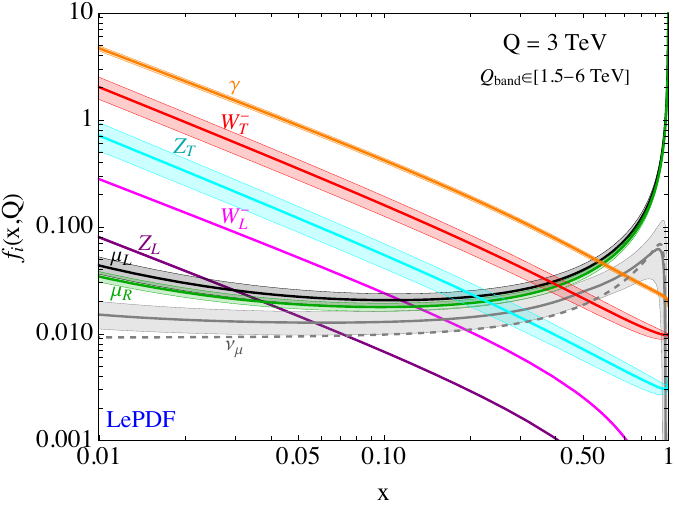}
    \caption{PDFs of a muon for a factorization scale $Q = 500\, \GeV$ (left) and $Q = 3\, \TeV$ (right), obtained with LePDFs \cite{Garosi:2023bvq}. The uncertainty bands correspond to the envelope obtained changing the factorization scale by a factor of $1/2$ and 2. The dashed gray line is the result for the muon neutrino PDF obtained with the $\mathcal{O}(\alpha)$ expression in Eq.~\eqref{eq:fnumu_NLO}.}
    \label{fig:LePDF}
\end{figure}

In the case of high-energy lepton colliders, specifically multi-TeV muon colliders, given the very large energy available in the hard scattering and the fact that electroweak radiation plays a leading role in the phenomenology (since leptons are not colored), it has been shown that complete EW interactions should be implemented in the PDFs and their evolution 
\cite{Chen:2016wkt,Dawson:1984gx,Kane:1984bb,Ciafaloni:2001mu,Ciafaloni:2005fm,Bauer:2017isx,Bauer:2017bnh,Bauer:2018arx,Han:2020uid,Han:2021kes,Garosi:2023bvq,Ciafaloni:2024alq}
In this work we use the recent implementation of LePDF \cite{Garosi:2023bvq}, which resums at leading-logarithmic order the complete set of SM DGLAP equations. We report some examples of PDFs in \cref{fig:LePDF}, see also Refs.~\cite{Han:2020uid,Han:2021kes} for other implementations.

Charged-current EW interaction has the peculiar property, not present in QED nor QCD, that it changes the fermion species: the analogous of the $\mu^- \to \mu^- \gamma$ splitting is $\mu^-_L \to \nu_\mu W^-$. As a consequence, already at $\mathcal{O}(\alpha_2)$ a muon neutrino PDF is induced.
While this splitting has the same IR soft divergence as in the QED case, EW loop corrections do not cancel it since the corresponding virtual correction only affects the muon PDF and not the neutrino one (see Sec.~3.2 of Ref.~\cite{Garosi:2023bvq} for a simple demonstration).
This is a violation of the Bloch-Nordsieck theorem \cite{Bloch:1937pw}, due to the initial and final states not being EW singlets. 
The IR divergence of the splitting therefore remains and is cutoff only by the $W$ boson mass. A consequence of this fact is the presence of Sudakov double logarithms, i.e. contributions to the PDFs that scale as $\alpha_2 \log^2 Q^2 / m_W^2$ \cite{Amati:1980ch,Ciafaloni:1998xg,Ciafaloni:2000gm,Ciafaloni:2000df,Ciafaloni:2000rp,Ciafaloni:2001vt,Manohar:2018kfx}.
The soft IR divergence also implies that most collinear gauge bosons are emitted with small energies and therefore we expect the muon neutrino PDF to increase when $x$ nears 1. This features can indeed be observed in \cref{fig:LePDF}. The subsequent fall of the muon neutrino PDF very close to $x=1$ is instead due to the infrared cutoff set by the $W$ mass.

\begin{figure}[t]
\centering
    \includegraphics[width=0.48\textwidth]{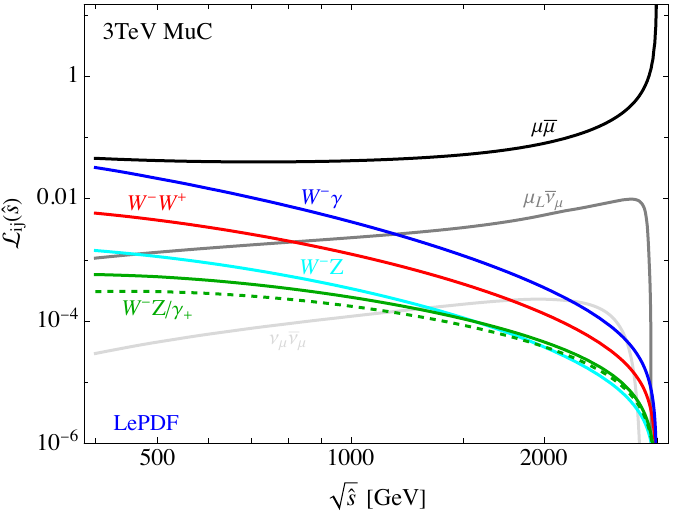} 
    \includegraphics[width=0.48\textwidth]{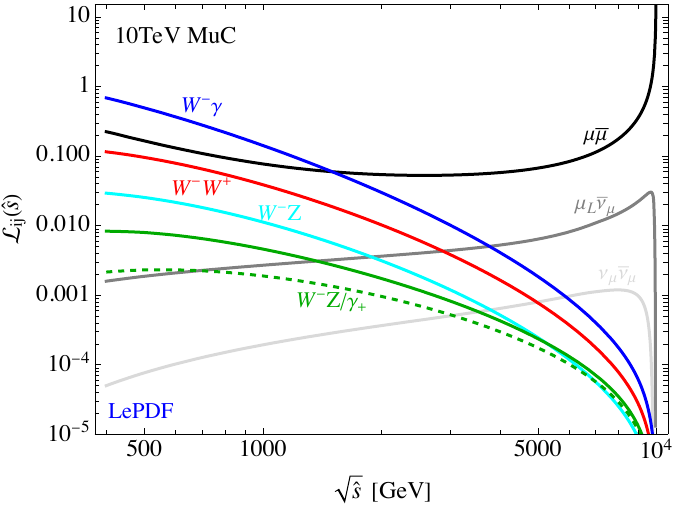} 
    \caption{Some examples of parton luminosities for a 3 TeV (left) and 10 TeV (right) MuC. The factorization scale is chosen as $Q = \sqrt{\hat{s}}/2$. Unless written explicitly, different polarizations are summed. For the $W^-$-$Z/\gamma$ luminosity we sum over the $W$ polarization and show with a solid (dashed) green line the absolute value of the positive (negative) $Z/\gamma$ helicity.}
    \label{fig:LePDF_lumi}
\end{figure}

As shown in Ref.~\cite{Garosi:2023bvq}, by solving the EW DGLAP equations iteratively at $\mathcal{O}(\alpha_2)$ we can derive an approximate analytic expression for the neutrino PDF: 
\be\begin{split}
    f_{\nu_\mu}^{(\alpha_2)}(x,Q^2) &= \frac{\alpha_2(Q)}{8\pi} \; \theta\!\left(Q^2 -\frac{m_W^2}{(1-x)^2}\right)\left[ \frac{1+x^2}{1-x} \left( \log \frac{Q^2 + x m_W^2}{m_W^2} +  \right. \right. \\
    & \qquad \left. + \log \frac{(1-x)^2}{1 + x(1-x)^2} + \frac{x m_W^2}{Q^2 + x m_W^2} + \frac{1}{1+x (1-x)^2} - 1\right) \\
   &\qquad \left. +\frac{2x^2(1-x)^2}{(1-x)(1+x(1-x)^2)}\frac{Q^2-m_W^2}{Q^2+xm_W^2}\right]~,
    \label{eq:fnumu_NLO}
\end{split}\ee
where $\theta$ is the Heaviside step function that follows from the IR cutoff mentioned above. The last line is due to ultra-collinear emission of a longitudinal $W$ boson from the muon. This analytic result is shown as a dashed gray line in \cref{fig:LePDF}, showing a good agreement with the full numerical result from LePDF. Further contributions to the neutrino PDF are expected at $\mathcal{O}(\alpha^2_2)$ mainly via the splitting $Z \to \bar{\nu} \nu$ and indeed we observe that the deviation grows at small $x$ and for larger factorization scales, where the $Z$ PDF is larger.

An estimate of the PDF uncertainties due to missing higher orders is typically obtained by varying the factorization scale. In \cref{fig:LePDF} we show with colored bands the envelope, for each PDF, obtained by varying the factorization scale $Q$ by a factor of 1/2 and 2 around the central value of 500 GeV (left) or 3 TeV (right). We see that these uncertainties are small for the photon, muon and the longitudinal polarization of EW bosons, while they are much larger for the transverse polarizations of $W$ and $Z$, and for the muon neutrino. This can be understood as follows. 
The longitudinal $W_L$ and $Z_L$ PDFs receive the dominant contribution from ultra-collinear emission off the valence muon. Such terms have no logarithmic scaling with $Q$ and approach instead a constant value at large scales \cite{Chen:2016wkt}, hence the very small scale dependence. 
The different scale dependence between the photon and muon on the one hand, and $W_T$, $Z_T$, and muon neutrino on the other, is due to the fact that the former have a leading contribution from QED interactions starting from the $m_\mu$ scale, i.e. scaling as $\log Q^2 / m_\mu^2$, while the latter evolve approximately as $\log Q^2/m_W^2$. For instance a variation of $Q$ by a factor of 2 around 500 GeV gives a relative $\sim 8\%$ effect in the case of QED contributions, compared to a $\sim 38\%$ change for EW ones. These reduce to $\sim 7\%$ and $\sim 19\%$, respectively, around $Q = 3 \TeV$.
It is clear that in order to obtain precise SM predictions such uncertainties should be reduced by deriving higher-order EW PDFs. Some discussions on possible extensions to higher order resummation can be found in Refs.~\cite{Manohar:2018kfx,Bauer:2018arx}. 

In \cref{fig:LePDF_lumi} we show some examples of parton luminosities (see Eq.~\eqref{eq:lumi}),
where for simplicity we sum over polarizations, except for $W^-$-$Z/\gamma$, where we sum only over the $W$ polarizations\footnote{The two $W^-$-$Z/\gamma_\pm$ luminosities have opposite sign and similar magnitude, giving an unphysical cancellation once they are summed.}. We fix $Q = \sqrt{\hat{s}}/2$ as factorization scale and do not report uncertainty bands, in order to not overcrowd the figure. It can be observed that for large invariant masses the $\mu^- \mu^+$ and $\mu_L^- \bar\nu_\mu$ luminosities dominate over the gauge bosons ones, as a consequence of the growth of both muon and neutrino PDFs for $x\to 1$.

%-------------------------------------------------
%-------------------------------------------------
\section{Assessing the $\nu_\mu$ PDF at muon colliders}
\label{sec:measurement}

In this Section, we investigate SM processes that are particularly sensitive to the neutrino PDF.
Specifically, we focus on $\BFmu \BFmubar \to e^- \bar\nu_e$ and $\BFmu \BFmubar \to W^- \gamma$, where the contribution from the $\nu_\mu$ PDF plays a leading role. These examples allow us to assess quantitatively the potential for testing experimentally the related SM predictions and could be used in the future to establish a proper treatment of EW radiation effects in high-energy processes.

%-------------------------------------------------
\subsection{Single-electron production}
\label{sec:monoe}

The main process which could be used as a probe of the muon neutrino PDF is $\BFmu \BFmubar \to e^- \bar\nu_e$, where the partonic process we are interested in is 
\begin{equation}
    \mu^-\bar{\nu}_\mu\to e^-\bar{\nu}_e, \label{eq:signal_enu}
\end{equation}
which proceeds via $s$-channel $W$ exchange. We refer to this as the signal.
The same final state can also be produced via VBF, which is our background:
\begin{equation}
\begin{split}
&W^-\gamma\to e^{-}\bar{\nu}_e,\\
&W^-Z\to e^{-}\bar{\nu}_e.\\
\end{split}\label{eq:bg_enu}
\end{equation}
\begin{figure}[t]
    \centering
    \includegraphics[width=0.8\textwidth]{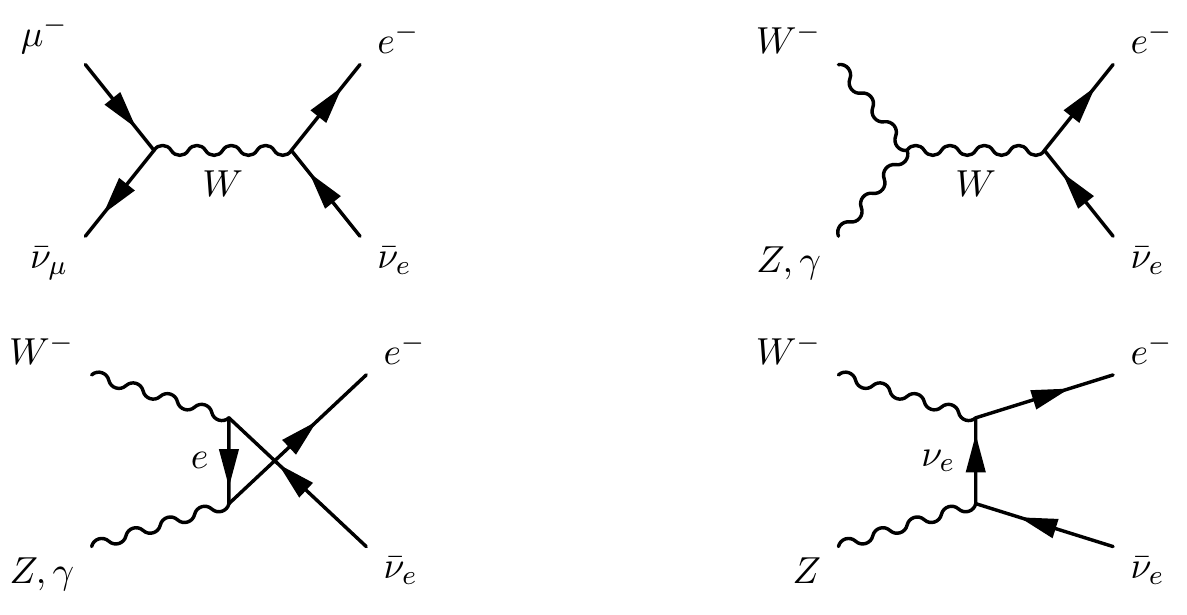}
    \caption{Leading partonic diagrams, in unitary gauge, contributing to $e^- \bar\nu_e$ production at a MuC.}
    \label{fig:diags_enu}
\end{figure}
The Feynman diagrams for these processes are depicted in \cref{fig:diags_enu}, where the upper left diagram corresponds to the signal and the others to the background.
The calculation of the background cross section includes the effect of interference between the photon and transverse $Z$ processes, which is then convoluted with the mixed $Z/\gamma$ PDF \cite{Ciafaloni:2000gm,Ciafaloni:2005fm,Chen:2016wkt,Garosi:2023bvq,Marzocca:2024fqb}.\footnote{A dedicated study on the effects due to to the $Z/\gamma$ PDF at MuC can be found in Ref.~\cite{Marzocca:2024fqb}.}
We neglect other background processes such as $e^- \bar{\nu}_e \to e^- \bar{\nu}_e$, $\tau^- \bar{\nu}_\tau \to e^- \bar{\nu}_e$, and $d_i \bar{u}_j \to e^- \bar{\nu}_e$, since their effects are further suppressed by small parton luminosities, see Ref.~\cite{Garosi:2023bvq}.

We can start to understand the relative weights of signal and background processes with simple estimates. The partonic cross sections of the two processes, for partonic invariant masses above the EW scale $\hat{s} \gg m_W$, follow the same scaling $\hat{\sigma}_{\mu\bar\nu} \sim \hat{\sigma}_{\rm VBF} \sim \alpha_{\rm EW}^2 / \hat{s}$. The physical cross section is obtained by convoluting these with the corresponding parton luminosities in \cref{fig:LePDF_lumi}. Since the $\mu \bar{\nu}_\mu$ luminosity dominates over the $VV$ ones at large invariant masses, we can expect that the signal will dominate the cross section in the high-energy region. This makes the $e^- \bar{\nu}_e$ process particularly sensitive to the neutrino PDF.

\begin{figure}[t]
    \centering
    \includegraphics[width=0.47\textwidth]{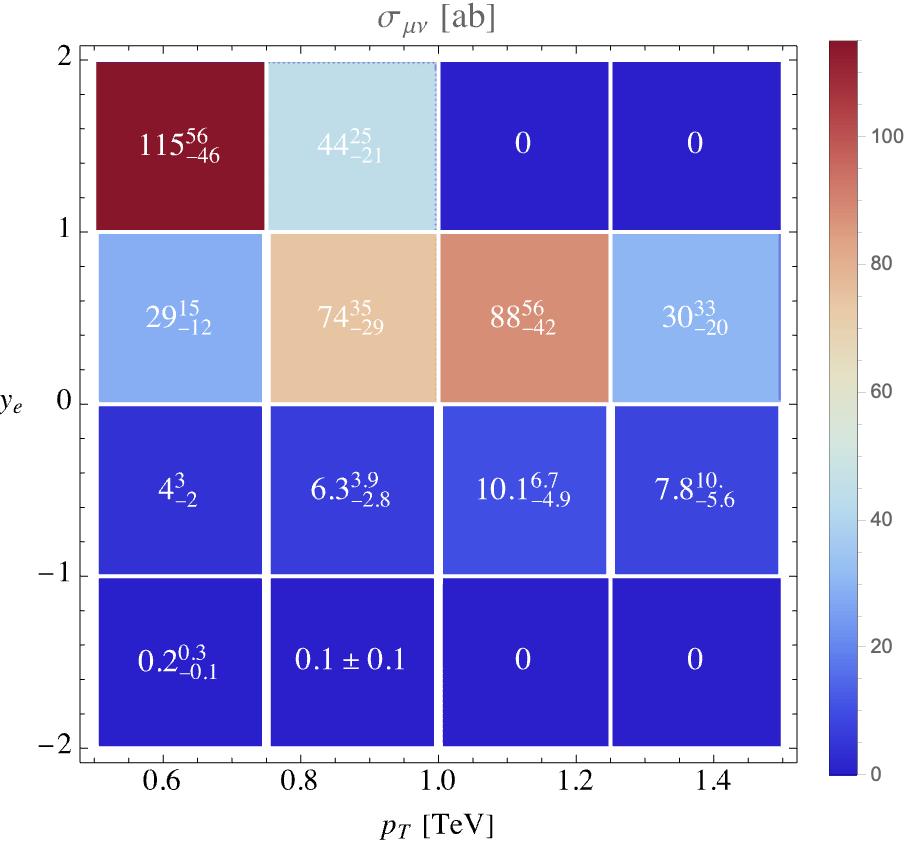}
    \includegraphics[width=0.47\textwidth]{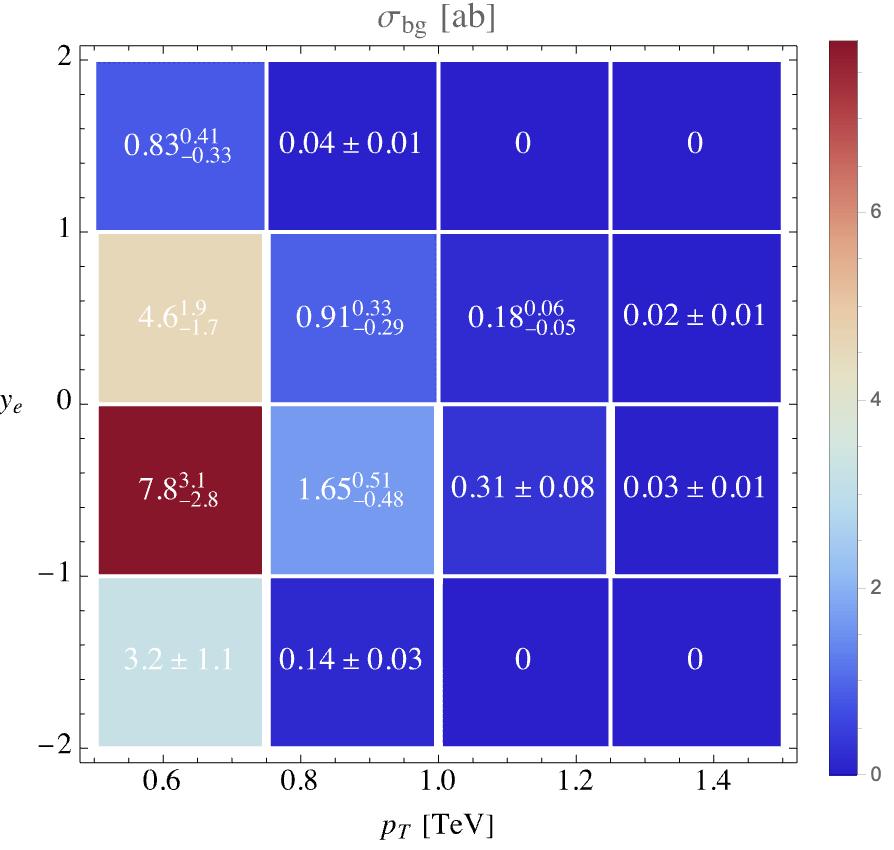}
    \includegraphics[width=0.47\textwidth]{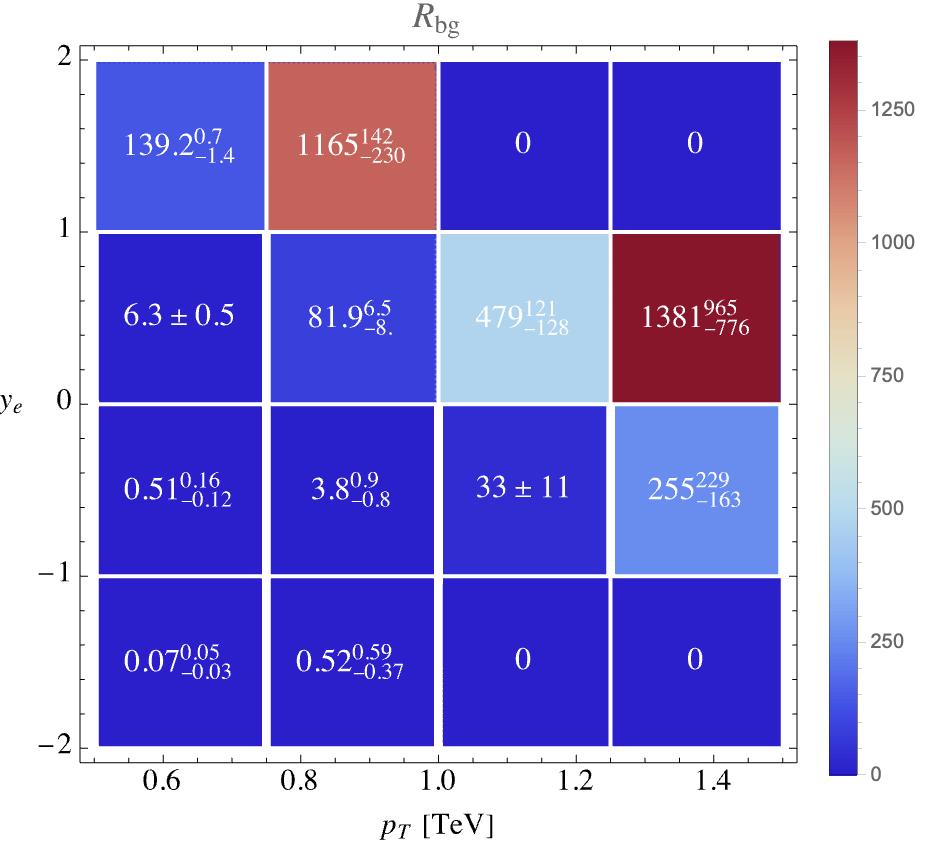}
    \caption{Top: Binned cross section for $\BFmu \BFmubar \to e^- \bar\nu_e$ at a 3 TeV Muon Collider, showing only the signal (left panel) or the background cross section (right panel).
    Bottom: Ratio of the signal cross section over the the VBF background.}
    \label{fig:bin_plot_enu}
\end{figure}

The physical triple differential cross sections for both signal and background are obtained by convoluting the partonic ones with the PDFs of the initial-state partons:\footnote{The formula is exact when the two partons and the final states are all massless. The generalization to massive partons, which is the case for the background, is straightforward and we checked that differences are negligible in the kinematical regime where collinear factorization can be applied, i.e. $p_T \gg m_W$.}
\begin{equation}
    \frac{d^3 \sigma(\BFmu \BFmubar \to e^- \bar\nu_e + X)}{d y_e d y_\nu d p_T} = \sum_{i,j} f_{i}^{\BFmu}\left(x_1,\frac{\hat{s}}{4}\right) f_{j}^{\BFmubar}\left(x_2,\frac{\hat{s}}{4}\right)  \left( \frac{2 p_T \hat{s}}{s_0} \right) \frac{d \hat{\sigma}}{d \hat{t}}(i j \to e^- \bar\nu_e)(\hat{s}, \hat{t})~,
    \label{eq:convolution_d3sigma_enu}
\end{equation}
We refer to Appendix~\ref{app:cross_section} for details and the expressions of the kinematical variables in terms of the final state's rapidities $y_{e,\nu}$ and $p_T$.
Since the neutrino cannot be detected, we integrate over $y_\nu$ to obtain the double differential cross section in terms of $y_e$ and $p_T$. We then bin these two variables to derive the total cross sections for both the signal and the background in each bin. Our results are shown in \cref{fig:bin_plot_enu}. The top-left panel displays the signal cross section (denoted by $\sigma_{\mu\nu}$), while the top-right panel shows the background cross section (denoted by $\sigma_{\rm bg}$). These cross sections correspond to a 3 TeV muon collider. We restrict the $p_T$ of the electron to values greater than 500 GeV to ensure the validity of the collinear approximation for electroweak PDFs, where $m_{\rm EW} \ll E_{\rm hard} \sim p_T^e$. The rapidity is considered within the interval [-2, 2], motivated by the geometric acceptance of the detector, ensuring that we focus on events away from the forward nozzles \cite{DiBenedetto:2018cpy}, which are generally cleaner and less affected by beam-induced backgrounds \cite{Mokhov:2011zzd}.

Comparing the two top panels in \cref{fig:bin_plot_enu}, we observe a significant difference between the signal and background distributions as a function of $p_T$. 
At relatively low $p_T\in(500, 750)$ GeV, the signal is largely forward, with its rapidity distribution peaking in the most forward bin ($y_e > 1$) and rapidly decreasing at lower rapidities ($y_e < 1$). In contrast, the background distribution peaks in the near-central region ($-1 < y_e < 0$) and is highly suppressed in the most forward rapidity bin. This behavior can be understood because the signal originates from a purely left-handed scattering amplitude, where the left-handed nature of the neutrino forces the final state to favor a helicity configuration that drives the electron forward. On the other hand, the background consists of vector bosons, whose helicity states can be either transverse or longitudinal. These states can add up to several helicity configurations, which contribute in different amounts to the total amplitude.

As the $p_T$ of the final state electron increases, both the signal and background distributions shift towards more central rapidities, due to kinematical constraints, see Eq.~\eqref{eq:y3_pt_constraints}. However, the background is significantly suppressed compared to the signal. This behavior arises from the fundamental difference between the PDFs of gauge bosons, which are peaked at small momentum fractions of the muon beam, and the ones of the muon and the muon neutrino, which instead are peaked at high momentum fractions, as discussed in \cref{sec:nuPDF} (see \cref{fig:LePDF}). To quantify the impact of neutrino PDF on this process, we define the following ratio
\begin{equation}
    R_{\rm bg}^{e\nu} = \frac{\sigma^{e\nu}_{\mu\nu}}{\sigma_{\rm bg}^{e\nu}},\label{eq:ratios_bins}
\end{equation}
where the label ${e\nu}$ is simply to specify the final state. This ratio can be used to establish the ideal place to look for neutrino PDF contributions, i.e. in bins where both the $R$ ratios and the cross section are big enough. The results are shown in the bottom panel of \cref{fig:bin_plot_enu}.
In conclusion, the contribution of the neutrino PDF to single-electron production is maximized at large $p_T$ and, for intermediate $p_T$ values, in the forward regions of the rapidity distribution.

%-------------------------------------------------
\subsection{$W \gamma$ production}
\label{ssec:Wgamma}

The associated production of a photon and a $W$, $\mu \bar{\mu} \to W^-\gamma$, follows the same lines as the single-electron production: at the parton level we have a neutrino-induced signal while the background proceeds via vector boson fusion:
\begin{equation}
\begin{split}
&\mu^-\bar{\nu}_\mu\to W^-\gamma,\\
&W^-\gamma\to W^-\gamma,\\
&W^-Z\to W^-\gamma.\\
\end{split} \label{eq:processes_wgamma}
\end{equation}

\begin{figure}[t]
    \centering
    \includegraphics[width=0.9\textwidth]{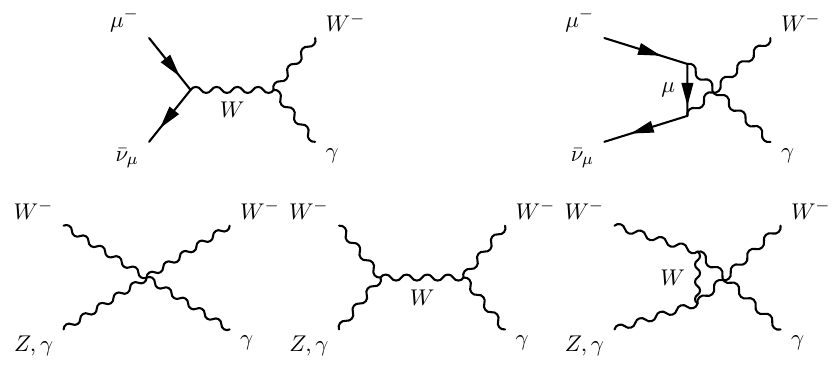}
    \caption{Leading partonic diagrams contributing to $W\gamma$ production at a MuC, in unitary gauge, for both the signal (upper line) and the background (lower line).}
    \label{fig:diags_wgamma}
\end{figure}

The relevant Feynman diagrams are shown in Fig.~\ref{fig:diags_wgamma}. As for single-electron production, we neglect background processes induced by initial-state electrons, taus, and quarks due to the suppression of parton luminosity.
It is worth noting that this process is less sensitive to the neutrino PDF, as the background partonic cross section $\hat{\sigma}_{\rm VBF}$ remains constant at high energies, whereas $\hat{\sigma}_{\mu\bar{\nu}} \sim \alpha_{\rm EW}^2/\hat{s}$, as previously discussed.

\begin{figure}[t]
    \centering
    \includegraphics[width=0.47\textwidth]{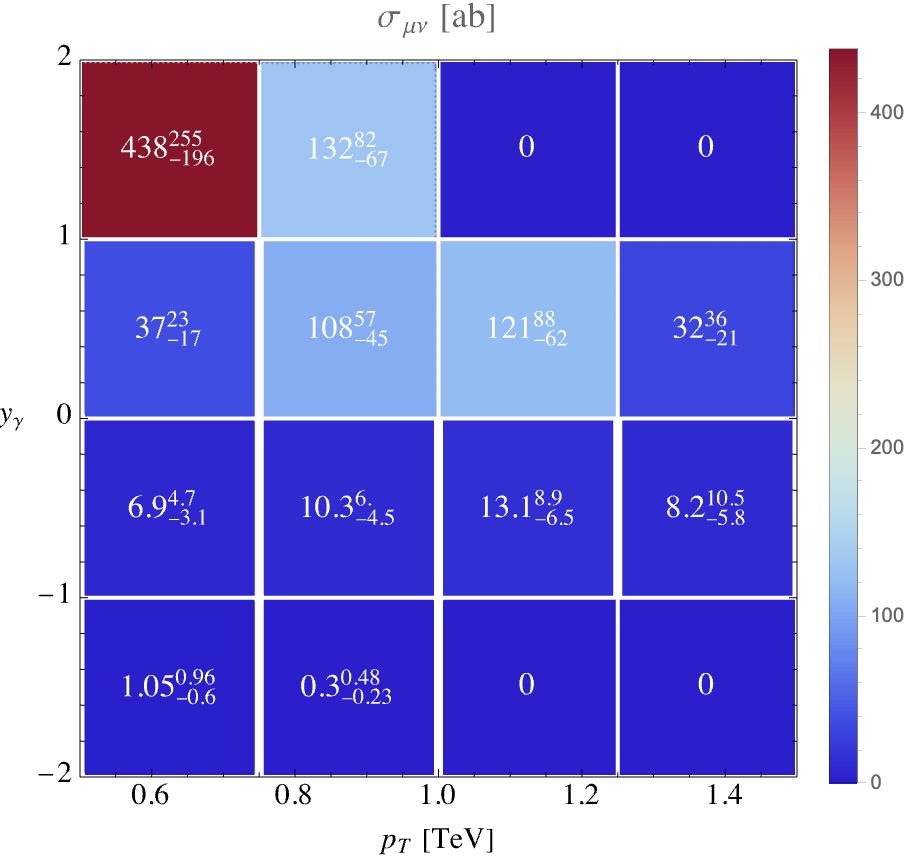}
    \includegraphics[width=0.47\textwidth]{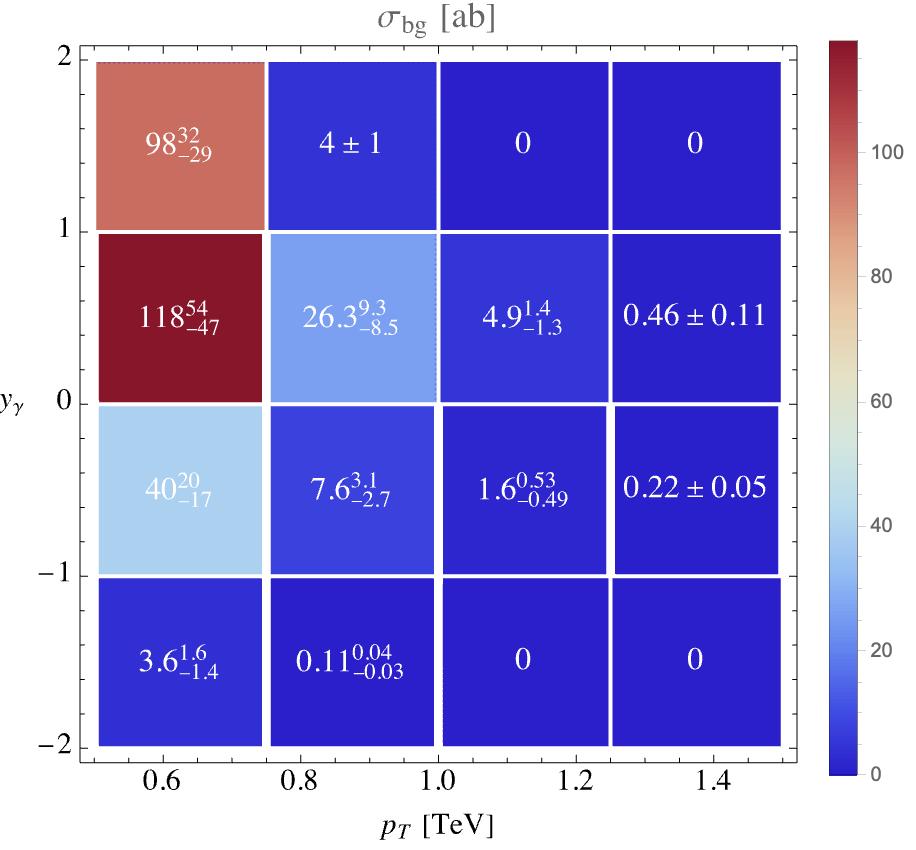}
    \includegraphics[width=0.47\textwidth]{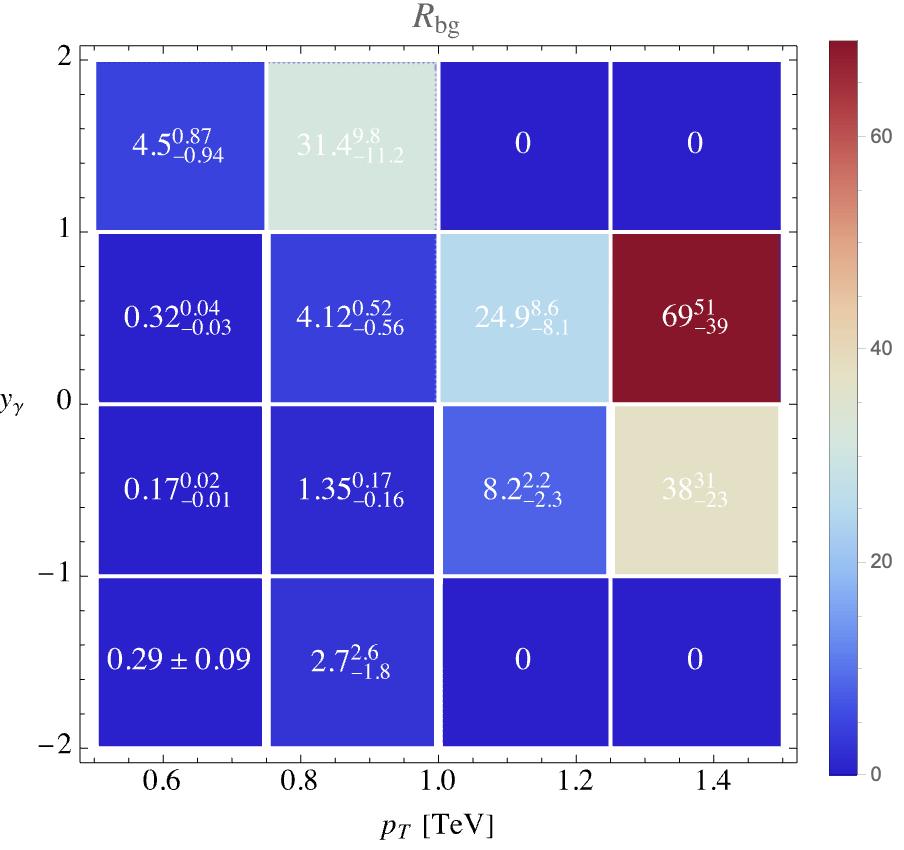}
    \caption{Top: Binned cross section for $\BFmu \BFmubar \to W^- \gamma$ at a 3 TeV Muon Collider, showing only the signal (left panel) or the background (right panel).
    Bottom: Ratio of the signal cross section over the VBF background.}
    \label{fig:bin_plot_wgamma}
\end{figure}

Since the rapidity of the $W$ boson is also measurable, one can, in principle, study triple-differential distributions for this process using the formula in Eq.~\eqref{eq:convolution_d3sigma}. Factorization holds when the invariant mass of the hard scattering is much larger than the electroweak scale, allowing us to neglect the particle masses in our analysis. For simplicity, we perform a similar analysis as in \cref{sec:monoe}, focusing on double-differential distributions in the transverse momentum ($p_T$) and the rapidity ($y_\gamma$) of the final-state photon.\footnote{We choose to use the photon rapidity being it easier to measure, without the need to reconstruct the $W$ from the decay products.} Our results for the signal ($\sigma_{\mu\nu}$) and background ($\sigma_{\rm bg}$) are shown in the top-left and top-right panels of \cref{fig:bin_plot_wgamma}, respectively. The signal is concentrated in the most forward rapidity region when $p_T$ is small, similar to the single-electron case. This behavior arises from the maximally broken parity of the initial-state fermions. In contrast to the single-electron case, however, the background is also strongly forward in the low-$p_T$ region. Nevertheless, as $p_T$ increases, the signal becomes more central, and the background is significantly suppressed, enabling a clearer measurement of the neutrino PDF contribution to $W\gamma$ production.

Finally, similar to the single-electron case, we define the ratio
\begin{equation} 
R_{\rm bg}^{W\gamma} = \frac{\sigma^{W\gamma}_{\mu\nu}}{\sigma_{\rm bg}^{W\gamma}}.
\label{eq:ratios_WA}
\end{equation}
The results are shown in the bottom panel of \cref{fig:bin_plot_wgamma}, which depict a very similar behavior to that of single-electron production.

An interesting aspect of the $W\gamma$ process, unlike the single-electron case, is that the momenta of both final-state particles can be reconstructed, allowing us to determine the center of mass of the $W\gamma$ system. This offers two key advantages: (1) we can compute the angular distribution of the $W$ boson relative to the boost of the system ($\cos\theta$), which reveals the so-called Radiation Amplitude Zero (RAZ) \cite{Brown:1979ux,Baur:1994sa,Baur:1994ia}.\footnote{The RAZ was first discovered in \cite{Mikaelian:1979nr}, observed at the Tevatron \cite{D0:2008abl}, and recently measured at the LHC \cite{CMS:2013ryd,CMS:2021cxr,ATLAS:2024qbd,ATLAS:2019bsc}. This feature of multi-boson scattering enhances sensitivity to new physics searches \cite{Capdevilla:2019zbx}.} For the $\mu\nu \to W^-\gamma$ process, the RAZ appears in the most forward region of the $\cos\theta$ distribution.
%\footnote{A detailed analysis of the RAZ and its detectability at multi-TeV MuC will be explored elsewhere \cite{Capdevilla:2024th}.}
(2) By reconstructing the center of mass energy, we can study how the $W\gamma$ cross section evolves with $\sqrt{\hat{s}}$. As $\sqrt{\hat{s}}$ increases, the background diminishes while the signal strengthens, mimicking the behavior of the luminosity functions in \cref{fig:LePDF_lumi}, as we already saw by quantifying how the neutrino PDF contribution dominates at high $p_T$.

%-------------------------------------------------
\subsection{Comparison between PDF and fixed-order approaches}
\label{sec:comparison}

As discussed above, employing PDFs to describe ISR gives a good approximation, and allows to resum leading logarithms, if the emitted radiation is collinear, that is its $p_T$ is much smaller than the typical energy scale involved in the hard scattering.
The main disadvantages of this approach is that the collinear radiation is integrated over, so cannot be described in a differential way, and that power-like corrections are neglected.

An alternative approach is to perform a fixed-order calculation, where the full process is described, including ISR. On the one hand, in this way the collinear radiation is fully described and can be used as an experimental handle to select interesting events, on the other hand the possibly large double logarithms are not resummed and therefore it might suffer from large uncertainties in the high-energy regime \cite{Ma:2024ayr}.
Our goal in this Section is to compare the two approaches, at the lowest order, in the case of the simple process of $e^- \bar{\nu}_e$ production. Specifically, in the following we focus only on the $\mu^- \bar{\nu}_\mu$ fusion contribution, neglecting VBF. As shown in \cref{sec:monoe}, the former is by far the dominant contribution in the hard region of large electron $p_T^e$.

\begin{figure}[t]
    \centering
    \includegraphics[width=0.9\textwidth]{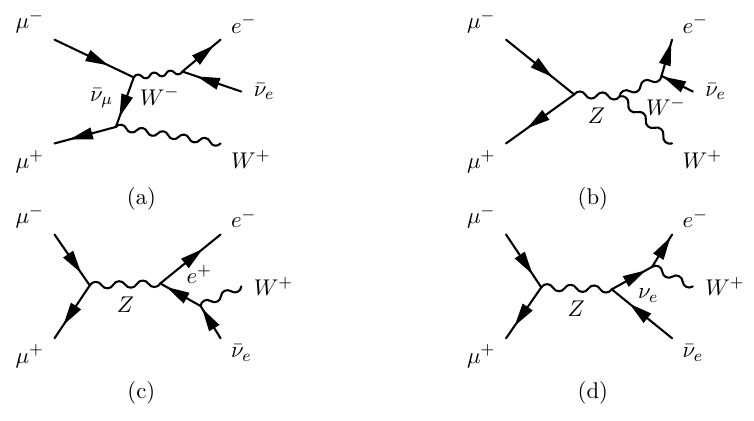}
    \caption{Leading-order Feynman diagrams for $\mu^- \mu^+ \to e^- \bar{\nu}_e W^+$.}
    \label{fig:diags_mumu_enuW}
\end{figure}

At leading order, single-electron production at muon colliders proceeds via $\mu^- \mu^+ \to e^- \bar{\nu}_e W^+$. The leading order Feynman diagrams are shown in \cref{fig:diags_mumu_enuW}. For the purpose of the comparison with the PDF approach we are interested in the region where the final-state $W$ is collinear and is emitted by the initial $\mu^+$, diagram (a).
This process however receives two other contributions, that should be removed with appropriate cuts to isolate the collinear $W$ emission. The first arises from on-shell $W^- W^+$ pair production, with $W^-$ decaying to electron-neutrino (\cref{fig:diags_mumu_enuW}-(b)). This contribution is characterized by an invariant mass of the electron-neutrino pair close to the $W$ mass and central $W$s. The second is through production of $e^- e^+$ or $\nu_e \bar{\nu}_e$ via neutral current, with subsequent emission, like final-state radiation (FSR), of a $W^+$ boson from either the $e^+$ or the $\nu_e$ (\cref{fig:diags_mumu_enuW}-(c,d)). These $W^+$ are typically emitted collinearly from the lepton and with small energies, due to the infrared singularity of the splitting function.

\begin{figure}[t]
    \centering
    \includegraphics[width=0.47\textwidth]{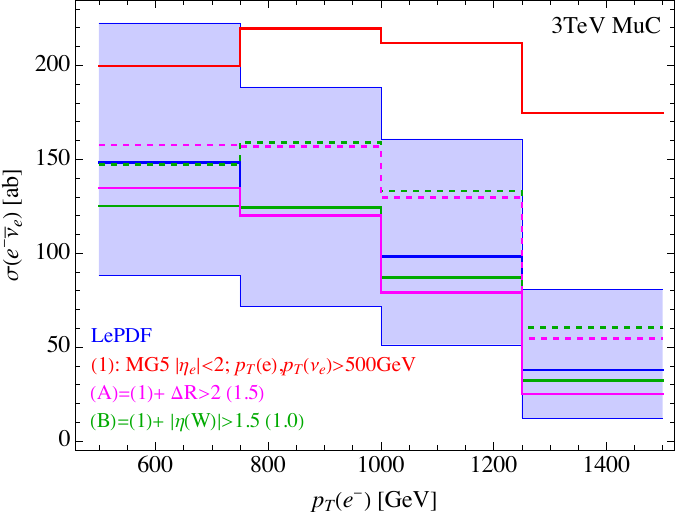}
    \includegraphics[width=0.47\textwidth]{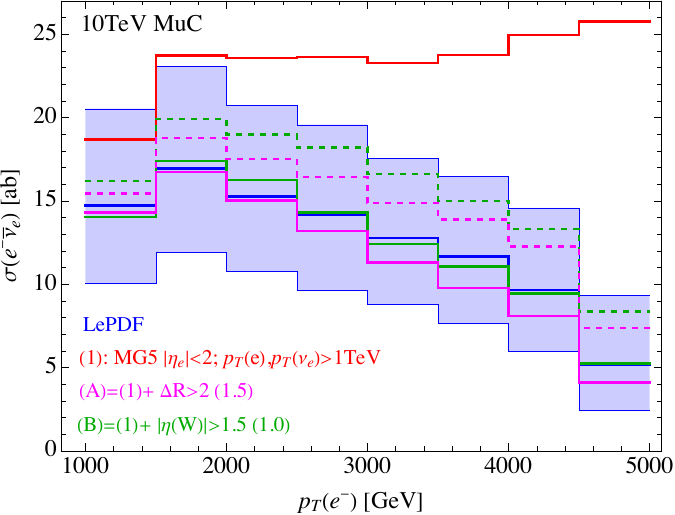}
    \caption{Comparison between PDF and fixed-order results for $e^- \bar\nu_e$ production at a 3 TeV (left) and 10 TeV (right) MuC, with subsequent cuts imposed on the events generated with {\tt MadGraph5\_aMC}. Dashed magenta and green lines correspond to the cuts shown in parentheses.}
    \label{fig:pTe_xsec}
\end{figure}

We study the cross section in different bins of $p_T^e$. For the PDF approach at a 3 TeV MuC we can use the result reported in \cref{fig:bin_plot_enu}, integrating over the electron rapidity $y_e$ between $-2$ and $2$. The corresponding cross section in $p_T^e$ bins, for both the 3 TeV and 10 TeV MuC are shown as a blue line in \cref{fig:pTe_xsec}, with the blue band representing the factorization scale uncertainty.
To evaluate the fixed-order cross section we generate with {\tt MadGraph5\_aMC} \cite{Alwall:2014hca} $\mu^- \mu^+ \to e^- \bar{\nu}_e W^+$ events at leading order with 3 and 10 TeV of total invariant mass.
Since our only goal is to compare the two theoretical approaches, in the following we pretend that the four momenta of all final-state particles can be fully reconstructed, including the neutrino.
At generation level we impose the following cuts, that we label as (1): 
\be\begin{split}
    (1)_{3 \TeV}: &~   |y_e|<2~, \quad p_T^e > 500 \,\GeV~, \quad p_T^{\nu} > 500 \,\GeV~, \quad M(e,\nu_e) > 200 \,\GeV~, \\
    (1)_{10 \TeV}: &~   |y_e|<2~, \quad p_T^e > 1 \,\TeV~, \quad p_T^{\nu} > 1 \,\TeV~, \quad M(e,\nu_e) > 500 \,\GeV~.
\end{split}\ee
The former two cuts are the same as the ones used in \cref{fig:bin_plot_enu} while the latter two, that are automatically satisfied in the PDF approach (since in that case $p_T^\nu = p_T^e$ and $M(e,\nu_e) = 2 p_T \cosh\left({(y_e-y_\nu)/2}\right) > 2 p_T$), are used to remove the otherwise overwhelming contribution from on-shell $WW$ production and to ensure the hardness of the $e^- \bar{\nu}_e$ system. We note that at this point almost all events have $M(e,\nu_e) \gtrsim 1 \, (2)\, \TeV$ for the 3 (10) TeV MuC and $\Delta \phi(e,\nu_e) \approx \pi$.
The resulting cross section in bins of $p_T^e$ is shown as a red line in \cref{fig:pTe_xsec}. One can notice a large mismatch with the result obtained from the PDF approach, that does not even improve with large $p_T^e$.
This is due to the $W^+$ emitted as FSR from final-state leptons, i.e. diagrams (c,d) in \cref{fig:diags_mumu_enuW}.

To remove this contribution and ensure that only the collinear $W$ region is selected, we present two alternatives. The first, that we denominate as (A), is to impose an isolation criteria on both the electron and the neutrino, requiring 
\be
    \text{(A)}: \quad \Delta R(i,j) > 2 \text{ or } 1.5~,
\ee
for all three final state particles $i,j = e^-, \bar\nu_e, W^+$, and $\Delta R = \sqrt{(\Delta \eta)^2 + (\Delta \phi)^2}$. While this is automatically satisfied for the electron-neutrino pair, given they have $\Delta \phi(e,\nu_e) \approx \pi$, this cut ensures that the $W^+$ is not collinear with the final-state leptons. The electron-$p_T$ distribution for this case is shown with magenta lines (respectively solid and dashed for the two values of the cut) in \cref{fig:pTe_xsec}, and shows a good agreement with the PDF result.

The second alternative, denominated as (B), is to impose a mild cut on the $W$ pseudo-rapidity,
\be
    \text{(B)}: \quad |\eta_W|>1.5 \text{ or } 1.0~.
\ee
Also this cut brings the fixed-order result in good agreement with the PDF one for both cut values, as shown with the green lines in \cref{fig:pTe_xsec} (respectively solid and dashed for the two cuts). 

\begin{figure}[t]
    \centering
    \includegraphics[width=0.46\textwidth]{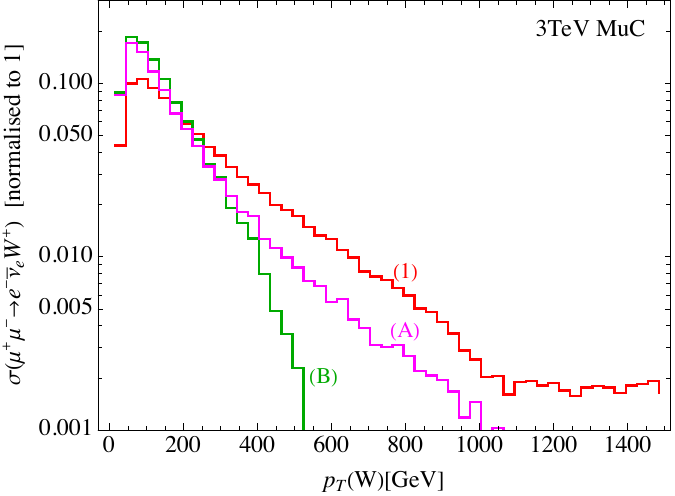}
    \includegraphics[width=0.47\textwidth]{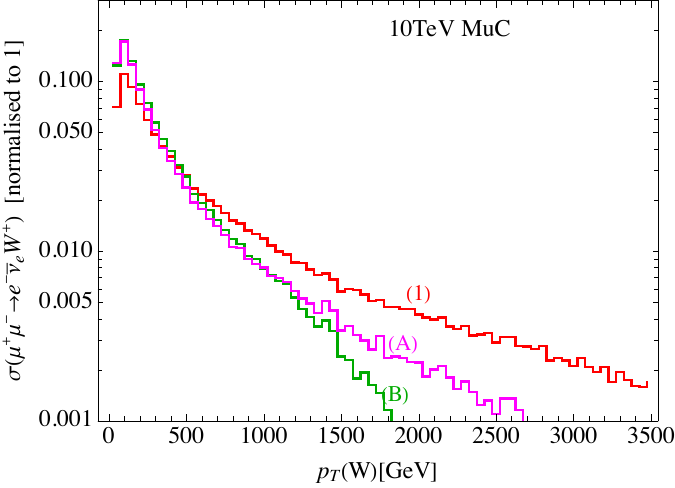}
    \caption{Normalized $p_T$ distribution of the $W$ boson for the various cut stages for the 3 and 10 TeV muon colliders.}
    \label{fig:enu_MG5}
\end{figure}

The $p_T$ distribution of the $W$ boson, normalized to 1, for each stage of the cuts described above is shown in \cref{fig:enu_MG5} and shows how the collinear condition is well satisfied once the cuts (A) or (B) are applied.

These results clearly demonstrate that the cuts used are effective in isolating the collinear $W$ emission and that the fixed-order calculations agree with the resummed predictions obtained from the PDF approach within the scale uncertainties.
However, there is a notable issue in this comparison: both cut types (A) and (B) are applied specifically to the $W$ boson, which contrasts with the PDF approach, where the treatment of collinear radiation is inclusive. This difference in methodology may lead to discrepancies when comparing results between the two approaches.
An indication of such discrepancies is presented by considering the difference between the solid and dashed lines, which corresponds to the two different values used in the cuts in (A) and (B).
While the cuts we use are likely broad enough to approximate inclusivity, a more rigorous comparison would require applying identical cuts to both the fixed-order and resummed approaches. This challenge highlights a broader issue in the treatment of electroweak radiation at high-energy muon colliders, a topic that extends beyond the scope of our current study.

%-------------------------------------------------
%-------------------------------------------------
\section{Charged-current pair production of heavy particles}
\label{sec:CCBSM}

The main production channel of heavy new physics states $X$ at muon colliders is through neutral-current pair production $\BFmu \BFmubar \to (\gamma/Z)^* \to X X^\dagger$, with a mass reach very close to the production threshold $E_{\rm MuC}/2$ \cite{Long:2020wfp,AlAli:2021let,Aime:2022flm,Black:2022cth,Accettura:2023ked,Andreetto2024,Accettura:2024qnk}, where $E_{\rm MuC}$ is the MuC center-of-mass energy. 
For a characterization of the new physics state it is however important to correlate this with results from other production channels. A possibility is to consider charged-current pair production, which for heavy masses would proceed via the muon neutrino component of the muon, since its PDF is peaked at large momentum fractions $x$: $\nu_{\mu} \mu^+ \to W^{+*} \to X_1 X_2^\dagger$, plus the conjugate process.
For concreteness, we consider a heavy scalar ${\rm SU}(2)_L$ doublet $S = (S_+, S_-)$ with hypercharge $Y=1/6$ (the main partonic process we are interested into does not depend on the hypercharge) and mass $M_S$.

\begin{figure}[t]
\centering
    \includegraphics[width=0.7\textwidth]{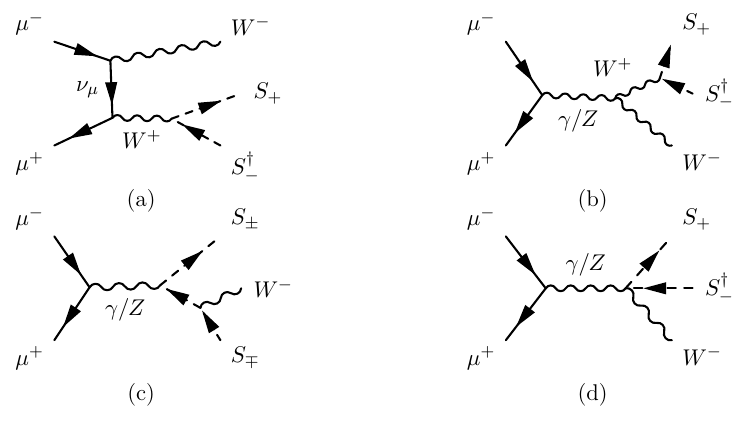} 
    \caption{Leading-order Feynman diagrams for $\mu^- \mu^+ \to S_+ S_-^\dagger W^-$.}\label{fig:diags_mumu_SSW}
\end{figure}

Our goal is to compare the total cross section, at a 10 TeV muon collider, derived by using LePDFs, relative to the process $\nu_{\mu} \mu^+ \to S_+ S_-^\dagger$ inclusive over collinear ISR, with the one obtained by a LO evaluation of $\mu^- \mu^+ \to S_+ S_-^\dagger W^-$, using MG5.
The fixed-order process receives contributions from several diagrams, reported in \cref{fig:diags_mumu_SSW}. In addition to the initial-state $W^-$-emission from the muon, diagram (a), there is also $W^+W^-$ production with photon or $Z$ in the $s$-channel (b), neutral-current production of $SS^\dagger$ with FSR emission of a $W^-$ from either of the two scalar legs, and contributions proportional to the quartic vertex $S_+ S_-^\dagger W^- \gamma (Z)$.
We expect that, for large $M_S$ values, the process of ISR collinear emission of a $W$ boson gives the dominant contribution, since it receives a $t$-channel enhancement when the intermediate neutrino has small $p_T$.

\begin{figure}[t]
    \centering
    \includegraphics[width=0.5\textwidth]{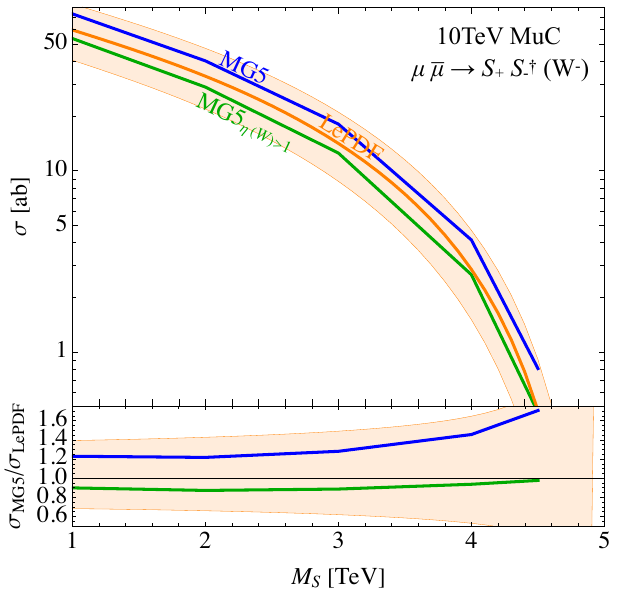}
    \caption{Cross section for the charged-current pair production of a heavy scalar doublet, at a 10 TeV MuC. Orange is the result obtained with LePDF (the band represents the scale uncertainty) with the muon neutrino PDF, blue and green are the {\tt MadGraph5\_aMC} results with no cuts and with a cut $\eta(W)>1$, respectively.}
    \label{fig:sc_doublet}
\end{figure}

The resulting cross sections, as function of the scalar mass $M_S$, are shown in \cref{fig:sc_doublet}.
We find a good agreement, within scale uncertainty, between the LePDF (orange, with uncertainty band obtained by varying the PDF factorization scale by a factor of 2) and MG5 (blue) results. 

\begin{figure}[t]
    \centering
    \includegraphics[width=0.32\textwidth]{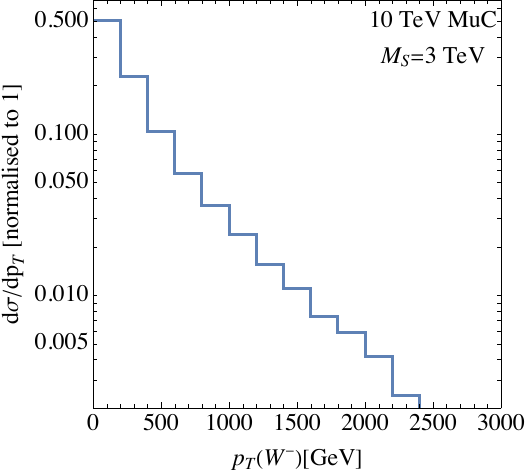}~
    \includegraphics[width=0.3\textwidth]{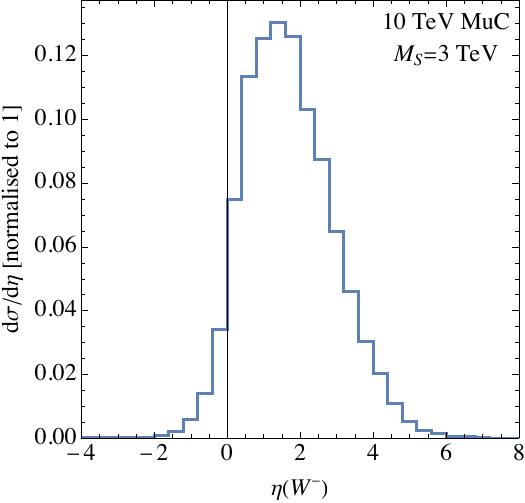}~
    \includegraphics[width=0.31\textwidth]{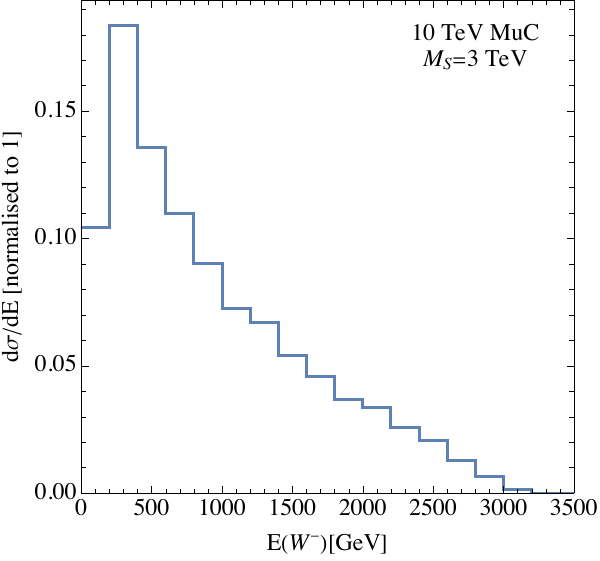}\\[5mm]
    \includegraphics[width=0.29\textwidth]{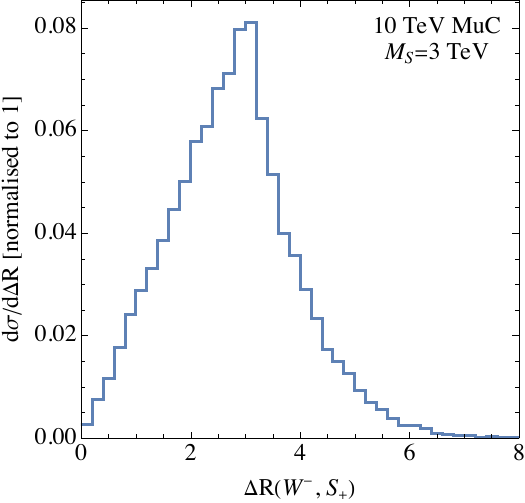}
    \includegraphics[width=0.3\textwidth]{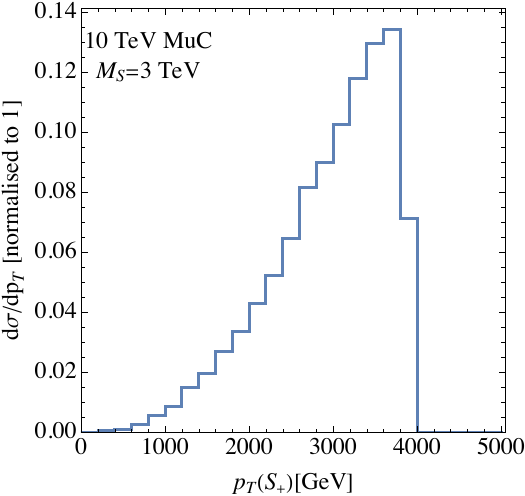}~
    \includegraphics[width=0.31\textwidth]{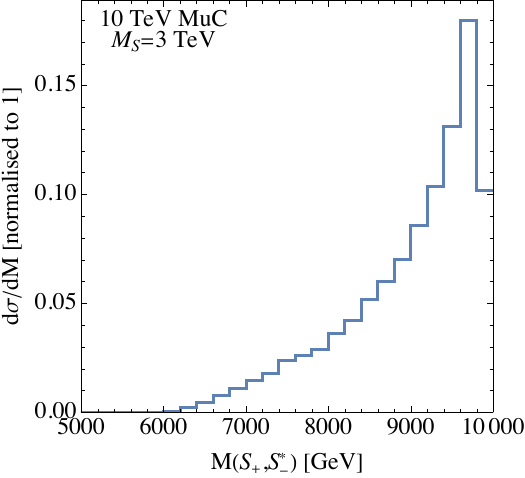}
    \caption{Kinematical distributions for the process $\mu^- \mu^+ \to S_+ S_-^\dagger W^-$ evaluated at LO with {\tt MadGraph5\_aMC}, for a 10 TeV MuC and $M_S = 3 \, \TeV$.}
    \label{fig:sc_doubl_MG}
\end{figure}

Some of the most relevant kinematical distributions for final state particles are shown in \cref{fig:sc_doubl_MG} for $M_S = 3\,\TeV$. 
We observe that the $W$ boson is indeed mostly forward and peaked at low values of $p_T$. The heavy scalars are instead produced centrally, with large $p_T$, and with very large invariant masses $M(S_+, S_-^\dagger)$.
In this case, contrary to the $e^- \bar\nu_e$ production studied above, the FSR contributions are not enhanced by large logarithms, since the state $S$ emitting the $W$ radiation is heavy, and indeed we observe a large angular separation, $\Delta R$, between $W$ and either of the $S_\pm$ scalars.
Nevertheless, to further suppress non-ISR contributions, which produce some central and even backward-going $W$ bosons, we impose a cut selecting only forward ones: $\eta(W^-)>1$. The resulting cross section is shown in green in \cref{fig:sc_doublet}, which agrees even better with the PDF result.

%

%-------------------------------------------------
%-------------------------------------------------
\section{Conclusions}
\label{sec:conclusions}

We have studied the neutrino content of a muon and the related phenomenology at a high energy muon collider. This specific aspect of EW PDFs falls into the broader topic of EW effects which become significant in the multi-TeV regime, offering new ways to test SM predictions in an unexplored energy range.

We reviewed how the muon neutrino PDF arises from the collinear emission of $W$ bosons from the muon and provided an analytic approximate expression for it, which is in good agreement with the result derived from the numerical leading-logarithmic resummation of SM DGLAP equations by LePDF.

We identified two SM processes, electron plus neutrino and $W\gamma$ production, that are particularly sensitive to this neutrino PDF, providing potential experimental handles to isolate and test the theoretical predictions. Our results indicate that both processes can serve as theoretical laboratories for high-precision tests of the muon neutrino PDF.
In case of electron plus neutrino production, we compared the results derived using the neutrino PDF with those obtained at leading-order from a Monte Carlo simulation. While our use of specific cuts in the fixed-order method was effective in isolating the collinear $W$ emission and in bringing the two results in agreement, this contrasts with the inclusive nature of the PDF approach, which accounts for all potential collinear emissions. Achieving a more rigorous comparison would require consistent treatment across both methods, a goal that shows the need for further theoretical developments in handling electroweak radiation at high-energy muon colliders.

We then applied the neutrino PDF to study the production of two heavy states via charged-current annihilation. This process could serve as an additional handle to understand the nature of the new physics states possibly discovered in neutral-current pair production.
As an example we took a scalar ${\rm SU(2)}_L$ doublet and computed the total cross section for charged-current pair production. We compared the results obtained by using LePDF with those of a leading-order simulation that includes explicitly the $W$ boson in the final state. In this case, contrary to the electron plus neutrino production, we find a good agreement even without imposing any cuts on the $W$. This is a consequence of the large masses of the scalars, which suppresses FSR $W$ emission compared to the emission from the initial state. Indeed, we observed that the $W$s are typically emitted collinearly and with small $p_T$.

Overall, our results underscore the importance of electroweak PDFs to both provide accurate SM predictions for muon collider processes and for new physics searches.
Further studies are required to improve these PDFs with higher-order corrections and to implement them in event-generation tools, both of which could be instrumental to better understand the complementarity between the resummed and fixed-order approaches to treat EW radiation. 
Addressing these challenges will be critical for maximizing the potential of future muon collider experiments in exploring physics beyond the Standard Model.

\section*{Acknowledgments}

The authors thank Davide Pagani and Andrea Wulzer for useful discussions.
We thank A. Wulzer also for pointing out an issue in the previous version of this work, regarding the applicability of the EW PDF formalism for single-Higgs production at muon colliders. That analysis has been removed in this version.
This manuscript has been authored by Fermi Research Alliance, LLC under Contract No.~DE-AC02-07CH11359 with the U.S. Department of Energy, Office of High Energy Physics.
DM acknowledges support by the Italian Ministry of University and Research (MUR) via the PRIN project n.~20224JR28W.

%%%%%%%%%%%%%%%%%%%%%%%%%%%%%%%%%%%
\appendix
%%%%%%%%%%%%%%%%%%%%%%%%%%%%%%%%%%%

%-------------------------------------------------
\section{Differential cross sections}
\label{app:cross_section}

For the production of two final state objects, assuming that the energy of the hard scattering is much larger than the $p_T$ of the initial-state collinear radiation and the masses of all objects, i.e. $p_T^{1,2}, \ m_{1,2,3,4} \ll p_T^{3,4}, \ \hat{s}$, then the triple-differential cross section for the production of a final state $X_3 X_4$ is given in terms of a convolution of the partonic cross section $d \hat{\sigma}/ d\hat{t}$ with the initial-state PDFs:
\begin{equation}
    \frac{d^3 \sigma(\BFmu \BFmubar \to X_3 X_4)}{d y_3 d y_4 d p_T} = \sum_{i,j} f_{i}^{\BFmu}\left(x_1,\frac{\hat{s}}{4}\right) f_{j}^{\BFmubar}\left(x_2,\frac{\hat{s}}{4}\right)  \left( \frac{2 p_T \hat{s}}{s_0} \right) \frac{d \hat{\sigma}}{d \hat{t}}(i j \to X_3 X_4)(\hat{s}, \hat{t})~,
    \label{eq:convolution_d3sigma}
\end{equation}
where $s_0$ is the collider center-of-mass energy squared, $\hat{s}$ and $\hat{t}$ are the partonic Mandelstam variables. All the kinematical variables can be expressed in terms of the two rapidities of the final-state particles, $y_{3,4}$, and their $p_T$ (by conservation of transverse momentum $p_T \equiv p_{T,3} = p_{T,4}$):
\begin{equation}\begin{split}
    y &\equiv \frac{y_3 - y_4}{2}~, \qquad Y \equiv \frac{y_3 + y_4}{2}~, \\
    \hat{s} &= 4 p_T^2 \cosh^2 y~, \qquad \hat{t} = - 2 p_T^2 e^{- y} \cosh y  ~, \\
    x_1 &= \frac{2p_T\cosh y}{\sqrt{s_0}} e^{Y}~, \qquad x_2 = \frac{2p_T\cosh y}{\sqrt{s_0}} e^{-Y}~.
\end{split}\end{equation}
Being $x_{1,2}\leq 1$, the kinematic constraints on $pT$, $y_3$ and $y_4$ are
\begin{equation}
    p_T\leq \frac{\sqrt{s_0}}{2} \frac{e^{\pm Y}}{\cosh y} \label{eq:pt_y_contraints}.
\end{equation}
Fixing $y_3$, the two functions at the RHS of Eq.~\eqref{eq:pt_y_contraints} intersect at $(y_4,p_T)=(-y_3,\sqrt{s_0}/2\cosh y_3)$. Then after integration on $y_4$ the following region is kinematically forbidden $\forall (y_3;p_T)$:
\begin{equation}
    p_T > \frac{\sqrt{s_0}}{2\cosh y_3}. \label{eq:y3_pt_constraints}
\end{equation}

%-------------------------------------------------
\subsection*{Parton luminosities}

A useful quantity that can be derived from PDFs are the parton luminosities, that describe the probability of having a certain pair of partons (one from each of the two colliding beams) with a given partonic invariant mass $\sqrt{\hat{s}}$, if the total invariant mass of the collision is $\sqrt{s_0}$. They are obtained as:
\be
    \mathcal{L}_{ij}(\hat{s}, s_0) = \int_0^1 \frac{dz}{z} f_{i;\BFmu}\left(z, \frac{\hat{s}}{4}\right) f_{j; \BFmubar} \left(\frac{\hat{s}}{z s_0}, \frac{\hat{s}}{4}\right)~.
    \label{eq:lumi}
\ee
Using parton luminosities, a total cross section for the production of a given hard final state $Y$, inclusive over collinear radiation $X$, can be computed simply by convoluting them with the partonic cross section:
\be
    \frac{d\sigma}{d \sqrt{\hat{s}}}(\BFmu \BFmubar \to Y + X) = \sum_{i,j} \frac{2 \sqrt{\hat s}}{s_0} \mathcal{L}_{ij}(\hat s, s_0) \hat\sigma(i j \to Y)(\hat s)~.
    \label{eq:dsigmadmlumi}
\ee
Depending on the situation it can be useful to derive the total cross section without using the parton luminosities but putting explicitly the PDFs:
\begin{equation}
    \label{eq:TotalCrossSection}
    \sigma(\BFmu \BFmubar \to Y + X)(s_0) = \sum_{i,j} \int_{0}^1 dx_1 \int_{0}^1 dx_2 \, f_i\left(x_1, \hat{s} / 4\right) \bar{f}_j\left(x_2, \hat{s}/4\right) \, \hat{\sigma}(i j \to Y)(\hat{s}),
\end{equation}
where the CoM energy of the interacting partons is given by $\hat{s} = x_1 x_2 s_0$.

%-------------------------------------------------
\vspace{1 cm}
\bibliographystyle{JHEP}
\bibliography{biblio}

\providecommand{\href}[2]{#2}\begingroup\raggedright\begin{thebibliography}{10}

\bibitem{Long:2020wfp}
K.~Long, D.~Lucchesi, M.~Palmer, N.~Pastrone, D.~Schulte and V.~Shiltsev, \emph{{Muon colliders to expand frontiers of particle physics}}, \href{https://doi.org/10.1038/s41567-020-01130-x}{\emph{Nature Phys.} {\bfseries 17} (2021) 289--292}, [\href{https://arxiv.org/abs/2007.15684}{{\ttfamily 2007.15684}}].

\bibitem{AlAli:2021let}
H.~Al~Ali et~al., \emph{{The muon Smasher\textquoteright{}s guide}}, \href{https://doi.org/10.1088/1361-6633/ac6678}{\emph{Rept. Prog. Phys.} {\bfseries 85} (2022) 084201}, [\href{https://arxiv.org/abs/2103.14043}{{\ttfamily 2103.14043}}].

\bibitem{Aime:2022flm}
C.~Aime et~al., \emph{{Muon Collider Physics Summary}},  \href{https://arxiv.org/abs/2203.07256}{{\ttfamily 2203.07256}}.

\bibitem{Black:2022cth}
K.~M. Black et~al., \emph{{Muon Collider Forum report}}, \href{https://doi.org/10.1088/1748-0221/19/02/T02015}{\emph{JINST} {\bfseries 19} (2024) T02015}, [\href{https://arxiv.org/abs/2209.01318}{{\ttfamily 2209.01318}}].

\bibitem{Accettura:2023ked}
C.~Accettura et~al., \emph{{Towards a muon collider}}, \href{https://doi.org/10.1140/epjc/s10052-023-11889-x}{\emph{Eur. Phys. J. C} {\bfseries 83} (2023) 864}, [\href{https://arxiv.org/abs/2303.08533}{{\ttfamily 2303.08533}}]. [Erratum: Eur.Phys.J.C 84, 36 (2024)].

\bibitem{Andreetto2024}
P.~Andreetto et~al., \emph{{Higgs Physics at a $\sqrt{s}=3$ TeV Muon Collider with detailed detector simulation}},  \href{https://arxiv.org/abs/2405.19314}{{\ttfamily 2405.19314}}.

\bibitem{Accettura:2024qnk}
C.~Accettura et~al., \emph{{Interim report for the International Muon Collider Collaboration (IMCC)}},  \href{https://arxiv.org/abs/2407.12450}{{\ttfamily 2407.12450}}.

\bibitem{Han:2020pif}
T.~Han, D.~Liu, I.~Low and X.~Wang, \emph{{Electroweak couplings of the Higgs boson at a multi-TeV muon collider}}, \href{https://doi.org/10.1103/PhysRevD.103.013002}{\emph{Phys. Rev. D} {\bfseries 103} (2021) 013002}, [\href{https://arxiv.org/abs/2008.12204}{{\ttfamily 2008.12204}}].

\bibitem{Forslund:2022xjq}
M.~Forslund and P.~Meade, \emph{{High precision higgs from high energy muon colliders}}, \href{https://doi.org/10.1007/JHEP08(2022)185}{\emph{JHEP} {\bfseries 08} (2022) 185}, [\href{https://arxiv.org/abs/2203.09425}{{\ttfamily 2203.09425}}].

\bibitem{Ruhdorfer:2023uea}
M.~Ruhdorfer, E.~Salvioni and A.~Wulzer, \emph{{Invisible Higgs boson decay from forward muons at a muon collider}}, \href{https://doi.org/10.1103/PhysRevD.107.095038}{\emph{Phys. Rev. D} {\bfseries 107} (2023) 095038}, [\href{https://arxiv.org/abs/2303.14202}{{\ttfamily 2303.14202}}].

\bibitem{Forslund:2023reu}
M.~Forslund and P.~Meade, \emph{{Precision Higgs width and couplings with a high energy muon collider}}, \href{https://doi.org/10.1007/JHEP01(2024)182}{\emph{JHEP} {\bfseries 01} (2024) 182}, [\href{https://arxiv.org/abs/2308.02633}{{\ttfamily 2308.02633}}].

\bibitem{Andreetto:2024rra}
P.~Andreetto et~al., \emph{{Higgs Physics at a $\sqrt{s}=3$ TeV Muon Collider with detailed detector simulation}},  \href{https://arxiv.org/abs/2405.19314}{{\ttfamily 2405.19314}}.

\bibitem{Li:2024joa}
P.~Li, Z.~Liu and K.-F. Lyu, \emph{{Higgs boson width and couplings at high energy muon colliders with forward muon detection}}, \href{https://doi.org/10.1103/PhysRevD.109.073009}{\emph{Phys. Rev. D} {\bfseries 109} (2024) 073009}, [\href{https://arxiv.org/abs/2401.08756}{{\ttfamily 2401.08756}}].

\bibitem{Han:2020uak}
T.~Han, Z.~Liu, L.-T. Wang and X.~Wang, \emph{{WIMPs at High Energy Muon Colliders}}, \href{https://doi.org/10.1103/PhysRevD.103.075004}{\emph{Phys. Rev. D} {\bfseries 103} (2021) 075004}, [\href{https://arxiv.org/abs/2009.11287}{{\ttfamily 2009.11287}}].

\bibitem{Buttazzo:2020uzc}
D.~Buttazzo, R.~Franceschini and A.~Wulzer, \emph{{Two Paths Towards Precision at a Very High Energy Lepton Collider}}, \href{https://doi.org/10.1007/JHEP05(2021)219}{\emph{JHEP} {\bfseries 05} (2021) 219}, [\href{https://arxiv.org/abs/2012.11555}{{\ttfamily 2012.11555}}].

\bibitem{Capdevilla:2021fmj}
R.~Capdevilla, F.~Meloni, R.~Simoniello and J.~Zurita, \emph{{Hunting wino and higgsino dark matter at the muon collider with disappearing tracks}}, \href{https://doi.org/10.1007/JHEP06(2021)133}{\emph{JHEP} {\bfseries 06} (2021) 133}, [\href{https://arxiv.org/abs/2102.11292}{{\ttfamily 2102.11292}}].

\bibitem{Bottaro:2021snn}
S.~Bottaro, D.~Buttazzo, M.~Costa, R.~Franceschini, P.~Panci, D.~Redigolo et~al., \emph{{Closing the window on WIMP Dark Matter}}, \href{https://doi.org/10.1140/epjc/s10052-021-09917-9}{\emph{Eur. Phys. J. C} {\bfseries 82} (2022) 31}, [\href{https://arxiv.org/abs/2107.09688}{{\ttfamily 2107.09688}}].

\bibitem{Chen:2022msz}
S.~Chen, A.~Glioti, R.~Rattazzi, L.~Ricci and A.~Wulzer, \emph{{Learning from radiation at a very high energy lepton collider}}, \href{https://doi.org/10.1007/JHEP05(2022)180}{\emph{JHEP} {\bfseries 05} (2022) 180}, [\href{https://arxiv.org/abs/2202.10509}{{\ttfamily 2202.10509}}].

\bibitem{Bottaro:2022one}
S.~Bottaro, D.~Buttazzo, M.~Costa, R.~Franceschini, P.~Panci, D.~Redigolo et~al., \emph{{The last complex WIMPs standing}}, \href{https://doi.org/10.1140/epjc/s10052-022-10918-5}{\emph{Eur. Phys. J. C} {\bfseries 82} (2022) 992}, [\href{https://arxiv.org/abs/2205.04486}{{\ttfamily 2205.04486}}].

\bibitem{Azatov:2022itm}
A.~Azatov, F.~Garosi, A.~Greljo, D.~Marzocca, J.~Salko and S.~Trifinopoulos, \emph{{New physics in b \textrightarrow{} s\ensuremath{\mu}\ensuremath{\mu}: FCC-hh or a muon collider?}}, \href{https://doi.org/10.1007/JHEP10(2022)149}{\emph{JHEP} {\bfseries 10} (2022) 149}, [\href{https://arxiv.org/abs/2205.13552}{{\ttfamily 2205.13552}}].

\bibitem{Liu:2023jta}
D.~Liu, L.-T. Wang and K.-P. Xie, \emph{{Composite resonances at a 10 TeV muon collider}}, \href{https://doi.org/10.1007/JHEP04(2024)084}{\emph{JHEP} {\bfseries 04} (2024) 084}, [\href{https://arxiv.org/abs/2312.09117}{{\ttfamily 2312.09117}}].

\bibitem{Korshynska:2024suh}
K.~Korshynska, M.~L\"oschner, M.~Marinichenko, K.~M\k{e}ka\l{}a and J.~Reuter, \emph{{Z\textquoteright{} boson mass reach and model discrimination at muon colliders}}, \href{https://doi.org/10.1140/epjc/s10052-024-12892-6}{\emph{Eur. Phys. J. C} {\bfseries 84} (2024) 568}, [\href{https://arxiv.org/abs/2402.18460}{{\ttfamily 2402.18460}}].

\bibitem{Capdevilla:2024bwt}
R.~Capdevilla, F.~Meloni and J.~Zurita, \emph{{Discovering Electroweak Interacting Dark Matter at Muon Colliders using Soft Tracks}},  \href{https://arxiv.org/abs/2405.08858}{{\ttfamily 2405.08858}}.

\bibitem{Amati:1980ch}
D.~Amati, A.~Bassetto, M.~Ciafaloni, G.~Marchesini and G.~Veneziano, \emph{{A Treatment of Hard Processes Sensitive to the Infrared Structure of QCD}}, \href{https://doi.org/10.1016/0550-3213(80)90012-7}{\emph{Nucl. Phys. B} {\bfseries 173} (1980) 429--455}.

\bibitem{Ciafaloni:1998xg}
P.~Ciafaloni and D.~Comelli, \emph{{Sudakov enhancement of electroweak corrections}}, \href{https://doi.org/10.1016/S0370-2693(98)01541-X}{\emph{Phys. Lett. B} {\bfseries 446} (1999) 278--284}, [\href{https://arxiv.org/abs/hep-ph/9809321}{{\ttfamily hep-ph/9809321}}].

\bibitem{Ciafaloni:2000gm}
M.~Ciafaloni, P.~Ciafaloni and D.~Comelli, \emph{{Electroweak double logarithms in inclusive observables for a generic initial state}}, \href{https://doi.org/10.1016/S0370-2693(01)00127-7}{\emph{Phys. Lett. B} {\bfseries 501} (2001) 216--222}, [\href{https://arxiv.org/abs/hep-ph/0007096}{{\ttfamily hep-ph/0007096}}].

\bibitem{Ciafaloni:2000df}
M.~Ciafaloni, P.~Ciafaloni and D.~Comelli, \emph{{Bloch-Nordsieck violating electroweak corrections to inclusive TeV scale hard processes}}, \href{https://doi.org/10.1103/PhysRevLett.84.4810}{\emph{Phys. Rev. Lett.} {\bfseries 84} (2000) 4810--4813}, [\href{https://arxiv.org/abs/hep-ph/0001142}{{\ttfamily hep-ph/0001142}}].

\bibitem{Ciafaloni:2000rp}
M.~Ciafaloni, P.~Ciafaloni and D.~Comelli, \emph{{Electroweak Bloch-Nordsieck violation at the TeV scale: 'Strong' weak interactions?}}, \href{https://doi.org/10.1016/S0550-3213(00)00508-3}{\emph{Nucl. Phys. B} {\bfseries 589} (2000) 359--380}, [\href{https://arxiv.org/abs/hep-ph/0004071}{{\ttfamily hep-ph/0004071}}].

\bibitem{Ciafaloni:2001vt}
M.~Ciafaloni, P.~Ciafaloni and D.~Comelli, \emph{{Bloch-Nordsieck violation in spontaneously broken Abelian theories}}, \href{https://doi.org/10.1103/PhysRevLett.87.211802}{\emph{Phys. Rev. Lett.} {\bfseries 87} (2001) 211802}, [\href{https://arxiv.org/abs/hep-ph/0103315}{{\ttfamily hep-ph/0103315}}].

\bibitem{Manohar:2018kfx}
A.~V. Manohar and W.~J. Waalewijn, \emph{{Electroweak Logarithms in Inclusive Cross Sections}}, \href{https://doi.org/10.1007/JHEP08(2018)137}{\emph{JHEP} {\bfseries 08} (2018) 137}, [\href{https://arxiv.org/abs/1802.08687}{{\ttfamily 1802.08687}}].

\bibitem{Chen:2016wkt}
J.~Chen, T.~Han and B.~Tweedie, \emph{{Electroweak Splitting Functions and High Energy Showering}}, \href{https://doi.org/10.1007/JHEP11(2017)093}{\emph{JHEP} {\bfseries 11} (2017) 093}, [\href{https://arxiv.org/abs/1611.00788}{{\ttfamily 1611.00788}}].

\bibitem{Dawson:1984gx}
S.~Dawson, \emph{{The Effective W Approximation}}, \href{https://doi.org/10.1016/0550-3213(85)90038-0}{\emph{Nucl. Phys. B} {\bfseries 249} (1985) 42--60}.

\bibitem{Kane:1984bb}
G.~L. Kane, W.~W. Repko and W.~B. Rolnick, \emph{{The Effective W+-, Z0 Approximation for High-Energy Collisions}}, \href{https://doi.org/10.1016/0370-2693(84)90105-9}{\emph{Phys. Lett. B} {\bfseries 148} (1984) 367--372}.

\bibitem{Ciafaloni:2001mu}
M.~Ciafaloni, P.~Ciafaloni and D.~Comelli, \emph{{Towards collinear evolution equations in electroweak theory}}, \href{https://doi.org/10.1103/PhysRevLett.88.102001}{\emph{Phys. Rev. Lett.} {\bfseries 88} (2002) 102001}, [\href{https://arxiv.org/abs/hep-ph/0111109}{{\ttfamily hep-ph/0111109}}].

\bibitem{Ciafaloni:2005fm}
P.~Ciafaloni and D.~Comelli, \emph{{Electroweak evolution equations}}, \href{https://doi.org/10.1088/1126-6708/2005/11/022}{\emph{JHEP} {\bfseries 11} (2005) 022}, [\href{https://arxiv.org/abs/hep-ph/0505047}{{\ttfamily hep-ph/0505047}}].

\bibitem{Bauer:2017isx}
C.~W. Bauer, N.~Ferland and B.~R. Webber, \emph{{Standard Model Parton Distributions at Very High Energies}}, \href{https://doi.org/10.1007/JHEP08(2017)036}{\emph{JHEP} {\bfseries 08} (2017) 036}, [\href{https://arxiv.org/abs/1703.08562}{{\ttfamily 1703.08562}}].

\bibitem{Bauer:2017bnh}
C.~W. Bauer, N.~Ferland and B.~R. Webber, \emph{{Combining initial-state resummation with fixed-order calculations of electroweak corrections}}, \href{https://doi.org/10.1007/JHEP04(2018)125}{\emph{JHEP} {\bfseries 04} (2018) 125}, [\href{https://arxiv.org/abs/1712.07147}{{\ttfamily 1712.07147}}].

\bibitem{Bauer:2018arx}
C.~W. Bauer and B.~R. Webber, \emph{{Polarization Effects in Standard Model Parton Distributions at Very High Energies}}, \href{https://doi.org/10.1007/JHEP03(2019)013}{\emph{JHEP} {\bfseries 03} (2019) 013}, [\href{https://arxiv.org/abs/1808.08831}{{\ttfamily 1808.08831}}].

\bibitem{Han:2020uid}
T.~Han, Y.~Ma and K.~Xie, \emph{{High energy leptonic collisions and electroweak parton distribution functions}}, \href{https://doi.org/10.1103/PhysRevD.103.L031301}{\emph{Phys. Rev. D} {\bfseries 103} (2021) L031301}, [\href{https://arxiv.org/abs/2007.14300}{{\ttfamily 2007.14300}}].

\bibitem{Han:2021kes}
T.~Han, Y.~Ma and K.~Xie, \emph{{Quark and gluon contents of a lepton at high energies}}, \href{https://doi.org/10.1007/JHEP02(2022)154}{\emph{JHEP} {\bfseries 02} (2022) 154}, [\href{https://arxiv.org/abs/2103.09844}{{\ttfamily 2103.09844}}].

\bibitem{Garosi:2023bvq}
F.~Garosi, D.~Marzocca and S.~Trifinopoulos, \emph{{LePDF: Standard Model PDFs for high-energy lepton colliders}}, \href{https://doi.org/10.1007/JHEP09(2023)107}{\emph{JHEP} {\bfseries 09} (2023) 107}, [\href{https://arxiv.org/abs/2303.16964}{{\ttfamily 2303.16964}}].

\bibitem{Ciafaloni:2024alq}
P.~Ciafaloni, G.~Co', D.~Colferai and D.~Comelli, \emph{{Electroweak evolution equations and isospin conservation}}, \href{https://doi.org/10.1007/JHEP07(2024)237}{\emph{JHEP} {\bfseries 07} (2024) 237}, [\href{https://arxiv.org/abs/2403.08583}{{\ttfamily 2403.08583}}].

\bibitem{Fermi:1924tc}
E.~Fermi, \emph{{On the Theory of the impact between atoms and electrically charged particles}}, \href{https://doi.org/10.1007/BF03184853}{\emph{Z. Phys.} {\bfseries 29} (1924) 315--327}.

\bibitem{Landau:1934zj}
L.~D. Landau and E.~M. Lifschitz, \emph{{ON THE PRODUCTION OF ELECTRONS AND POSITRONS BY A COLLISION OF TWO PARTICLES}}, \href{https://doi.org/10.1016/B978-0-08-010586-4.50021-3}{\emph{Phys. Z. Sowjetunion} {\bfseries 6} (1934) 244}.

\bibitem{vonWeizsacker:1934nji}
C.~F. von Weizsacker, \emph{{Radiation emitted in collisions of very fast electrons}}, \href{https://doi.org/10.1007/BF01333110}{\emph{Z. Phys.} {\bfseries 88} (1934) 612--625}.

\bibitem{Williams:1934ad}
E.~J. Williams, \emph{{Nature of the high-energy particles of penetrating radiation and status of ionization and radiation formulae}}, \href{https://doi.org/10.1103/PhysRev.45.729}{\emph{Phys. Rev.} {\bfseries 45} (1934) 729--730}.

\bibitem{Costantini:2020stv}
A.~Costantini, F.~De~Lillo, F.~Maltoni, L.~Mantani, O.~Mattelaer, R.~Ruiz et~al., \emph{{Vector boson fusion at multi-TeV muon colliders}}, \href{https://doi.org/10.1007/JHEP09(2020)080}{\emph{JHEP} {\bfseries 09} (2020) 080}, [\href{https://arxiv.org/abs/2005.10289}{{\ttfamily 2005.10289}}].

\bibitem{Ruiz:2021tdt}
R.~Ruiz, A.~Costantini, F.~Maltoni and O.~Mattelaer, \emph{{The Effective Vector Boson Approximation in high-energy muon collisions}}, \href{https://doi.org/10.1007/JHEP06(2022)114}{\emph{JHEP} {\bfseries 06} (2022) 114}, [\href{https://arxiv.org/abs/2111.02442}{{\ttfamily 2111.02442}}].

\bibitem{Alwall:2014hca}
J.~Alwall, R.~Frederix, S.~Frixione, V.~Hirschi, F.~Maltoni, O.~Mattelaer et~al., \emph{{The automated computation of tree-level and next-to-leading order differential cross sections, and their matching to parton shower simulations}}, \href{https://doi.org/10.1007/JHEP07(2014)079}{\emph{JHEP} {\bfseries 07} (2014) 079}, [\href{https://arxiv.org/abs/1405.0301}{{\ttfamily 1405.0301}}].

\bibitem{Kunszt:1987tk}
Z.~Kunszt and D.~E. Soper, \emph{{On the Validity of the Effective $W$ Approximation}}, \href{https://doi.org/10.1016/0550-3213(88)90673-6}{\emph{Nucl. Phys. B} {\bfseries 296} (1988) 253--289}.

\bibitem{Borel:2012by}
P.~Borel, R.~Franceschini, R.~Rattazzi and A.~Wulzer, \emph{{Probing the Scattering of Equivalent Electroweak Bosons}}, \href{https://doi.org/10.1007/JHEP06(2012)122}{\emph{JHEP} {\bfseries 06} (2012) 122}, [\href{https://arxiv.org/abs/1202.1904}{{\ttfamily 1202.1904}}].

\bibitem{Cuomo:2019siu}
G.~Cuomo, L.~Vecchi and A.~Wulzer, \emph{{Goldstone Equivalence and High Energy Electroweak Physics}}, \href{https://doi.org/10.21468/SciPostPhys.8.5.078}{\emph{SciPost Phys.} {\bfseries 8} (2020) 078}, [\href{https://arxiv.org/abs/1911.12366}{{\ttfamily 1911.12366}}].

\bibitem{Gribov:1972ri}
V.~N. Gribov and L.~N. Lipatov, \emph{{Deep inelastic e p scattering in perturbation theory}}, {\emph{Sov. J. Nucl. Phys.} {\bfseries 15} (1972) 438--450}.

\bibitem{Dokshitzer:1977sg}
Y.~L. Dokshitzer, \emph{{Calculation of the Structure Functions for Deep Inelastic Scattering and e+ e- Annihilation by Perturbation Theory in Quantum Chromodynamics.}}, {\emph{Sov. Phys. JETP} {\bfseries 46} (1977) 641--653}.

\bibitem{Altarelli:1977zs}
G.~Altarelli and G.~Parisi, \emph{{Asymptotic Freedom in Parton Language}}, \href{https://doi.org/10.1016/0550-3213(77)90384-4}{\emph{Nucl. Phys. B} {\bfseries 126} (1977) 298--318}.

\bibitem{Bloch:1937pw}
F.~Bloch and A.~Nordsieck, \emph{{Note on the Radiation Field of the electron}}, \href{https://doi.org/10.1103/PhysRev.52.54}{\emph{Phys. Rev.} {\bfseries 52} (1937) 54--59}.

\bibitem{Marzocca:2024fqb}
D.~Marzocca and A.~Stanzione, \emph{{On the impact of the mixed $Z/ \gamma$ PDF at muon colliders}},  \href{https://arxiv.org/abs/2408.13191}{{\ttfamily 2408.13191}}.

\bibitem{DiBenedetto:2018cpy}
V.~Di~Benedetto, C.~Gatto, A.~Mazzacane, N.~V. Mokhov, S.~I. Striganov and N.~K. Terentiev, \emph{{A Study of Muon Collider Background Rejection Criteria in Silicon Vertex and Tracker Detectors}}, \href{https://doi.org/10.1088/1748-0221/13/09/P09004}{\emph{JINST} {\bfseries 13} (2018) P09004}, [\href{https://arxiv.org/abs/1807.00074}{{\ttfamily 1807.00074}}].

\bibitem{Mokhov:2011zzd}
N.~V. Mokhov and S.~I. Striganov, \emph{{Detector Background at Muon Colliders}}, \href{https://doi.org/10.1016/j.phpro.2012.03.761}{\emph{Phys. Procedia} {\bfseries 37} (2012) 2015--2022}, [\href{https://arxiv.org/abs/1204.6721}{{\ttfamily 1204.6721}}].

\bibitem{Brown:1979ux}
R.~W. Brown, D.~Sahdev and K.~O. Mikaelian, \emph{{W+- Z0 and W+- gamma Pair Production in Neutrino e, p p, and anti-p p Collisions}}, \href{https://doi.org/10.1103/PhysRevD.20.1164}{\emph{Phys. Rev. D} {\bfseries 20} (1979) 1164}.

\bibitem{Baur:1994sa}
U.~Baur, S.~Errede and G.~L. Landsberg, \emph{{Rapidity correlations in $W \gamma$ production at hadron colliders}}, \href{https://doi.org/10.1103/PhysRevD.50.1917}{\emph{Phys. Rev. D} {\bfseries 50} (1994) 1917--1930}, [\href{https://arxiv.org/abs/hep-ph/9402282}{{\ttfamily hep-ph/9402282}}].

\bibitem{Baur:1994ia}
U.~Baur, T.~Han and J.~Ohnemus, \emph{{Amplitude zeros in W+- Z production}}, \href{https://doi.org/10.1103/PhysRevLett.72.3941}{\emph{Phys. Rev. Lett.} {\bfseries 72} (1994) 3941--3944}, [\href{https://arxiv.org/abs/hep-ph/9403248}{{\ttfamily hep-ph/9403248}}].

\bibitem{Mikaelian:1979nr}
K.~O. Mikaelian, M.~A. Samuel and D.~Sahdev, \emph{{The Magnetic Moment of Weak Bosons Produced in p p and p anti-p Collisions}}, \href{https://doi.org/10.1103/PhysRevLett.43.746}{\emph{Phys. Rev. Lett.} {\bfseries 43} (1979) 746}.

\bibitem{D0:2008abl}
{\scshape D0} collaboration, V.~M. Abazov et~al., \emph{{First study of the radiation-amplitude zero in $W \gamma$ production and limits on anomalous $W W \gamma$ couplings at $\sqrt{s}$ = 1.96- TeV}}, \href{https://doi.org/10.1103/PhysRevLett.100.241805}{\emph{Phys. Rev. Lett.} {\bfseries 100} (2008) 241805}, [\href{https://arxiv.org/abs/0803.0030}{{\ttfamily 0803.0030}}].

\bibitem{CMS:2013ryd}
{\scshape CMS} collaboration, S.~Chatrchyan et~al., \emph{{Measurement of the $W\gamma$ and $Z\gamma$ Inclusive Cross Sections in $pp$ Collisions at $\sqrt s=7$ TeV and Limits on Anomalous Triple Gauge Boson Couplings}}, \href{https://doi.org/10.1103/PhysRevD.89.092005}{\emph{Phys. Rev. D} {\bfseries 89} (2014) 092005}, [\href{https://arxiv.org/abs/1308.6832}{{\ttfamily 1308.6832}}].

\bibitem{CMS:2021cxr}
{\scshape CMS} collaboration, A.~Tumasyan et~al., \emph{{Measurement of W$^\pm\gamma$ differential cross sections in proton-proton collisions at $\sqrt{s}$ = 13 TeV and effective field theory constraints}}, \href{https://doi.org/10.1103/PhysRevD.105.052003}{\emph{Phys. Rev. D} {\bfseries 105} (2022) 052003}, [\href{https://arxiv.org/abs/2111.13948}{{\ttfamily 2111.13948}}].

\bibitem{ATLAS:2024qbd}
{\scshape ATLAS} collaboration, G.~Aad et~al., \emph{{Studies of the Energy Dependence of Diboson Polarization Fractions and the Radiation-Amplitude-Zero Effect in WZ Production with the ATLAS Detector}}, \href{https://doi.org/10.1103/PhysRevLett.133.101802}{\emph{Phys. Rev. Lett.} {\bfseries 133} (2024) 101802}, [\href{https://arxiv.org/abs/2402.16365}{{\ttfamily 2402.16365}}].

\bibitem{ATLAS:2019bsc}
{\scshape ATLAS} collaboration, M.~Aaboud et~al., \emph{{Measurement of $W^{\pm}Z$ production cross sections and gauge boson polarisation in $pp$ collisions at $\sqrt{s} = 13$ TeV with the ATLAS detector}}, \href{https://doi.org/10.1140/epjc/s10052-019-7027-6}{\emph{Eur. Phys. J. C} {\bfseries 79} (2019) 535}, [\href{https://arxiv.org/abs/1902.05759}{{\ttfamily 1902.05759}}].

\bibitem{Capdevilla:2019zbx}
R.~M. Capdevilla, R.~Harnik and A.~Martin, \emph{{The radiation valley and exotic resonances in $W\gamma$ production at the LHC}}, \href{https://doi.org/10.1007/JHEP03(2020)117}{\emph{JHEP} {\bfseries 03} (2020) 117}, [\href{https://arxiv.org/abs/1912.08234}{{\ttfamily 1912.08234}}].

\bibitem{Ma:2024ayr}
Y.~Ma, D.~Pagani and M.~Zaro, \emph{{EW corrections and Heavy Boson Radiation at a high-energy muon collider}},  \href{https://arxiv.org/abs/2409.09129}{{\ttfamily 2409.09129}}.

\end{thebibliography}\endgroup

%\end{fmffile}
\end{document}